\documentclass[traditabstract,usenatbib]{aa} 
\newcommand{\rmn}{\mathrm}
\usepackage{graphicx}
\usepackage{epstopdf}
\usepackage{natbib}
\usepackage{txfonts}
\usepackage{color}

\begin{document}

\title{High-energy gamma-ray and neutrino backgrounds from\\ clusters of
galaxies and radio constraints}

\author{
Fabio Zandanel\inst{1,\ast} \and
Irene Tamborra\inst{1} \and
Stefano Gabici\inst{2} \and
Shin'ichiro Ando\inst{1}
}
\institute{GRAPPA Institute, University of Amsterdam, Science Park 904, 1098 XH Amsterdam, Netherlands
\and APC, Univ.~Paris Diderot, CNRS/IN2P3, CEA/Irfu, Obs. de Paris, Sorbonne Paris Cit\'e, France\\
$\ast$~\email{f.zandanel@uva.nl}}


\abstract{
Cosmic-ray protons accumulate for cosmological times in clusters of galaxies because their typical radiative and diffusive escape times are longer than the 
Hubble time. Their hadronic interactions with protons of the intra-cluster medium generate secondary electrons, gamma rays, and neutrinos. 
In light of the high-energy neutrino events recently discovered by the IceCube neutrino observatory, for which galaxy clusters have been suggested as possible 
sources, and the forthcoming results from the {\it Fermi} gamma-ray survey, we here estimate the contribution from galaxy clusters to the diffuse gamma-ray 
and neutrino backgrounds. We modelled the cluster population by means of their mass function, using a phenomenological luminosity-mass relation 
applied to all clusters, as well as a detailed semi-analytical model. In the latter model, we divide clusters into cool-core/non-cool-core,
and loud/quiet subsamples, as suggested by observations, and model the cosmic-ray proton population according to state-of-the-art hydrodynamic numerical
simulations. Additionally, we consider observationally-motivated values for the cluster magnetic field. This is a crucial parameter since the observed radio counts of 
clusters need to be
respected owing to synchrotron emission by secondary electrons. For a choice of parameters respecting current constraints from radio to 
gamma   rays, and assuming a proton spectral index of $-2$, we find that hadronic interactions in clusters contribute less than 10\% to the IceCube flux
and much less to the total extragalactic gamma-ray background observed by {\it Fermi}. They account for less than 1\%  for spectral indices $\le-2$.
The high-energy neutrino flux observed by IceCube can be reproduced without violating radio constraints only if a very hard (and speculative) spectral 
index $>-2$ is adopted. However, this scenario is in tension with the high-energy IceCube  data, which seems to suggest a spectral energy distribution 
of the neutrino flux that decreases with the particle energy. We prove that IceCube should be able to test our most optimistic scenarios for spectral 
indices $\ge-2.2$ by stacking a few nearby massive galaxy clusters. In the case of proton-photon interactions in clusters, we find that very likely protons do not 
reach sufficiently high energies to produce neutrinos in these environments. We argue that our results are optimistic because of our assumptions and that 
clusters of galaxies cannot make any relevant contribution to the extragalactic gamma-ray and neutrino backgrounds in any realistic scenario.
Finally, we find that the cluster contribution to the angular fluctuations in the gamma-ray background is subdominant, less than 10\% on sub-degree scales.
}

\keywords{Galaxies: clusters: general -- Gamma rays: diffuse background -- Gamma rays: galaxies: clusters -- Neutrinos} 

\titlerunning{Gamma-ray and neutrino backgrounds from galaxy clusters}

\authorrunning{Zandanel et al.}

\maketitle

\section{Introduction}
\label{sec:1}
The extragalactic gamma-ray background (EGB) is the measured radiation that remains
after subtracting all known sources from the observed gamma-ray sky. The EGB was measured  by
the SAS-2 satellite for the first time~\citep{1977ApJ...217L...9F}  then by EGRET
\citep{1998ApJ...494..523S, 2004ApJ...613..956S} and the \emph{Fermi}-Large
Area Telescope (LAT; \citealp{2010PhRvL.104j1101A,2014arXiv1410.3696T}) most recently. The EGB is likely due to the 
sum of contributions from different unresolved sources, such as active galactic nuclei (AGN), star-forming 
galaxies, pulsars, gamma-ray bursts, and intergalactic shocks produced by structure formation~(see, e.g., \citealp{Dermer:2007fg,2010ApJ...720..435A,Stecker:2010di,2011MNRAS.415.1074S,
Ackermann:2012vca,2013MNRAS.429.1529F,2014ApJ...780..161D,
2014ApJ...786..129D,2014arXiv1404.1189T,2015arXiv150105301A,2015arXiv150105316D} and references therein). 

Recently, the IceCube neutrino observatory at the South Pole has reported evidence of 
extraterrestrial neutrinos~\citep{Aartsen:2013bka,Aartsen:2014gkd}.
The four-year IceCube dataset consists of $37$ events that exceed the
atmospheric background with a significance of more than $5 \sigma$~\citep{Aartsen:2014gkd}. 
The neutrino data are compatible with a flux isotropically distributed in the sky, with astrophysical origin 
and with a possible cutoff at a few PeV.  
The origin of these events is unknown (see \citealp{Waxman:2013zda} and
\citealp{Anchordoqui:2013dnh} for recent reviews; see also \citealp{2014PhRvD..90j3003W}). However, the
isotropic distribution in the sky of the observed events suggests that they might come from various extragalactic 
$\sim100$~PeV cosmic-ray (CR) accelerators, such as gamma-ray bursts, especially untriggered ones~\citep{1997PhRvL..78.2292W,2012PhRvL.108w1101H,Murase:2013ffa,Liu:2012pf};
AGN \citep{1999PhRvD..59b3002W,2005APh....23..537H,Stecker:2013fxa,2013PhRvD..88h3007W,Murase:2014foa,2014PhRvD..89l3005B}; star-forming
galaxies including starbursts, galaxy mergers, and
AGN~\citep{Loeb:2006tw,2014arXiv1404.1189T,Lacki:2010vs,
Murase:2013rfa,He:2013cqa,Liu:2013wia,Katz:2013ooa,Kashiyama:2014rza,Anchordoqui:2014yva,Chang:2014hua,2014arXiv1411.2783T};
intergalactic shocks and active galaxies embedded in structured
regions~\citep{Murase:2013rfa}; and hypernovae and supernova remnants~\citep{Chakraborty:2015sta,Senno:2015tra}. A galactic origin for the neutrinos has also been proposed~\citep{Ahlers:2013xia,Fox:2013oza,Joshi:2013aua,Taylor:2014hya,Anchordoqui:2014rca}, 
as well as mixed scenarios of galactic and extragalactic neutrino
sources~\citep{Ahlers:2013xia,Razzaque:2013uoa,Fox:2013oza,Joshi:2013aua,Murase:2014foa,Padovani:2014bha}. Exotic
models including PeV dark matter decay scenarios have  been 
discussed, too~\citep{Feldstein:2013kka,Esmaili:2013gha,Esmaili:2014rma}. 

As shown in~\cite{2013PhRvD..88l1301M}, a multi-messenger connection between the measured neutrino 
fluxes and their photon counterparts could be crucial for unveiling the origin of the
high-energy neutrinos, regardless of the physics 
of their sources. In the following, we assume  that the IceCube high-energy neutrinos have an extragalactic origin and are produced in proton-proton
collisions. In such a scenario we would expect sources to also emit gamma rays  at a flux comparable to that of neutrinos (see, e.g., \citealp{2006PhRvD..74c4018K}); however, the neutrinos could also be produced in proton-photon 
interactions (see, e.g., \citealp{2008PhRvD..78c4013K}).

Clusters of galaxies are the latest and largest structures to form in the Universe. During their 
assembly, energies of the same order of magnitude as the gravitational binding energy, $10^{61}$--$10^{63}$~erg, 
should be dissipated through structure-formation shocks and turbulence \citep{2005RvMP...77..207V}. 
Therefore, even if only a small part of this energy goes into particle acceleration, clusters should host 
significant non-thermal emission from radio to gamma rays (see, e.g., \citealp{2014IJMPD..2330007B}). 

The contribution of clusters of galaxies to the EGB has been discussed by several authors
\citep{2000Natur.405..156L,2003ApJ...585..128K,2003APh....19..679G,Ando:2007yw,2014MNRAS.440..663Z}. 
It has been argued that CR hadronic interactions in galaxy clusters could be responsible for a neutrino flux that is comparable 
to the one recently observed by IceCube \citep{2008ApJ...689L.105M,2009ApJ...707..370K,2013JCAP...02..028M,2013PhRvD..88l1301M}.
However, such hadronic interactions  could have a dramatic impact on the radio frequencies
since secondary electrons are also produced in proton-proton interactions and radiate synchrotron 
emission when interacting with the magnetic fields in clusters of galaxies. The radio emission from secondary 
electrons needs to respect radio counts of galaxy clusters~\citep{1999NewA....4..141G,VenturiGMRT_1,VenturiGMRT_2,2013arXiv1306.3102K}, 
since the cluster diffuse synchrotron radio emission has been observed (see, e.g., \citealp{2012A&ARv..20...54F}). 

In this work, we estimate the possible contribution to the extragalactic gamma-ray and neutrino backgrounds
from galaxy clusters assuming that gamma rays and neutrinos mainly originate in proton-proton interactions, 
while for the first time taking the consequences in the radio regime into account. 
We compare our model estimates to the isotropic diffuse gamma-ray background 
measured by {\it Fermi}  \citep{2014arXiv1410.3696T} and to the neutrino flux measured by
IceCube~\citep{Aartsen:2014gkd}.
We also discuss the small-scale anisotropies in EGB recently detected
with {\it Fermi} \citep{2012PhRvD..85h3007A} and compare the measurements
with cluster models.

This paper is organised as follows. In Section~\ref{sec:2}, we briefly discuss proton-proton
interactions in galaxy clusters and explain how we calculate the emission from secondary
electrons, photons, and neutrinos. We then introduce the mass function of galaxy clusters 
and a phenomenological luminosity-mass relation in Section~\ref{sec:3}.
In Section~\ref{sec:4}, we refine our approach by using a detailed semi-analytical model
based on state-of-the-art numerical simulations of CRs in clusters and test the robustness 
of our results with respect to the adopted parameters. We compare our results with
stacking upper limits by the IceCube telescope  and discuss future detection prospects
in Section~\ref{sec:4a}. We briefly discuss the neutrino 
contribution from proton-photon interactions in clusters in Section~\ref{sec:5} 
and the angular power spectrum of the EGB in Section~\ref{sec:6}. Finally, in Section~\ref{sec:7},
we summarise our findings.

\section{Secondaries from proton-proton interactions}
\label{sec:2}
The CR protons accumulate in galaxy clusters for cosmological times \citep{1996SSRv...75..279V,1997ApJ...487..529B} 
and interact with the thermal protons of the intra-cluster medium (ICM) generating secondary particles: electrons, neutrinos,
and high-energy photons \citep{1980ApJ...239L..93D,1999APh....12..169B,2001ApJ...559...59M,2004A&A...413...17P,2007IJMPA..22..681B,2008MNRAS.385.1211P,
2009JCAP...08..002K,2009ApJ...707..370K,2010MNRAS.409..449P}. 
While the ICM density is typically well known from X-ray measurements of its bremsstrahlung emission, 
the CR proton spectral and spatial distributions in galaxy clusters are unknown. In fact, whereas the diffuse radio emission observed in 
several clusters proves the presence of relativistic electrons, direct proof of proton acceleration has yet to be found.

Gamma-ray observations of the possible hadronic-induced emission started to put tight constraints on the proton content of clusters
\citep{2009arXiv0907.0727T,2009A&A...495...27A,2010ApJ...710..634A,2010ApJ...717L..71A,2011arXiv1111.5544M,2012...VERITAS,2012JCAP...07..017A,2013arXiv1308.6278H,
2013arXiv1310.5707V,2014MNRAS.440..663Z,2013arXiv1308.5654T,2013arXiv1309.0197P,2014arXiv1405.7047G}. Gamma-ray limits also suggest that secondary electrons cannot 
be uniquely responsible for the observed radio emission in galaxy clusters, at least in the case of the so-called giant radio haloes found in merging 
clusters like Coma \citep{2012arXiv1207.3025B,2012arXiv1207.6410Z}. As we discuss in the following, an important implication for our 
purposes is that the observed radio counts represent an optimistic upper limit for the radio emission from secondary electrons since only a 
fraction of it can have a hadronic origin.

Assuming a power law in momentum for the spectral distribution of CR
protons in clusters, $f(p) dp=\rho_{\rmn{CR}} p^{-\alpha_{\rmn{p}}} dp$, 
the radio synchrotron luminosity of secondary electrons at a frequency $f$ can be expressed as (adapted from \citealp{2008MNRAS.385.1211P})
\begin{equation}
L_{f} = A_{f} \int  \rho_{\rmn{CR}} \, \rho_{\rmn{ICM}} \frac{\epsilon_{B}}{\epsilon_{B}+\epsilon_{\rmn{CMB}}} \left( \frac{\epsilon_{B}}{\epsilon_{B_\rmn{c}}}\right)^{\frac{\alpha_{\rmn{p}-2}}{4}}  \,\rmn{d}V\ ,
\label{eq:Lradio}
\end{equation}
where $\rho_{\rmn{CR}}$ and $\rho_{\rmn{ICM}}$ are the CR proton and ICM density distributions, respectively, while $\epsilon_{B} = B^{2}/8\pi$ 
and $\epsilon_{\rmn{CMB}}$ are the energy densities of the cluster
magnetic fields and the cosmic microwave background (CMB)\footnote{The
total energy density of photons should also include  the contribution from
star light: $\epsilon_{{ph}} = \epsilon_{\rmn{stars}} +
\epsilon_{\rmn{CMB}}$.
However,  $\epsilon_{\rmn{stars}}$ is subdominant in the cluster volume
(see, e.g., Figure~5 of \citealp{2011arXiv1105.3240P}), therefore
$\epsilon_{{ph}} \approx \epsilon_{\rmn{CMB}}$.}. The
parameter $\epsilon_{B_\rmn{c}}$ is the magnetic energy density
corresponding to a characteristic magnetic field $B_{\rmn{c}} =31
(\nu/\rmn{GHz})$~$\mu$G for the synchrotron mechanism, and $A_{f}$
encloses the spectral information \citep{2008MNRAS.385.1211P}.
The gamma-ray luminosity of secondary photons at an energy $E$ is
defined as
\begin{equation}
L_{\gamma} = A_{\gamma} \int  \rho_{\rmn{CR}} \, \rho_{\rmn{ICM}} \,  \,\rmn{d}V\ ,
\label{eq:Lgamma}
\end{equation}
with $A_{\gamma}$ enclosing the spectral information \citep{2008MNRAS.385.1211P}.

In the following, we make use of Equations~(\ref{eq:Lradio}) and (\ref{eq:Lgamma}) to calculate the hadronic-induced emission 
in galaxy clusters at radio and gamma-ray frequencies. The spectral multipliers $A_{f}$ and $A_{\gamma}$ were obtained 
in \cite{2004A&A...413...17P} as analytical approximations of full proton-proton interaction simulations.  The 
analytical expressions for $A_{f}$ and $A_{\gamma}$ reproduce the results of numerical simulations from 
energies around the pion bump ($\sim$100~MeV) up to a few hundred GeV.  A more precise formalism
has been derived by \cite{2006PhRvD..74c4018K} for the TeV--PeV energy range, relevant to calculating the neutrino fluxes. Therefore, 
 we correct the gamma-ray spectra obtained by adopting the analytical
 approximations with the recipe in~\cite{2006PhRvD..74c4018K} for
 energies above $\sim$0.1--1~TeV. The transition energy between the two
 approximations depends on $\alpha_{\rm p}$, and it was chosen as the energy at which the two models coincide.

We compute the corresponding neutrino spectra as prescribed in~\cite{2006PhRvD..74c4018K}. 
When assuming that proton-proton interactions are the main interactions producing neutrinos and gamma rays, 
the neutrino intensity for all flavours could also be approximately obtained as a function of  the gamma-ray flux~\citep{Ahlers:2013xia,Anchordoqui:2004eu}: $L_{\nu} (E_{\nu}) \approx 6 \, L_{\gamma} (E_{\gamma})$, 
with $E_{\nu} \approx E_{\gamma}/2$,  where we ignored the absorption
during the propagation of gamma rays for simplicity. From this approximation, one finds that, at a given energy, $L_{\nu}/L_{\gamma} \sim1.5$ for $\alpha_{\rmn{p}} = 2$.
However, detailed calculations by \cite{1997ApJ...487..529B} and \cite{2006PhRvD..74c4018K} show that this ratio is slightly smaller
for spectral indices $\alpha_{\rmn{p}} > 2$ and slightly higher for $\alpha_{\rmn{p}} < 2$.

We do not assume any CR spectral cut-off at high energies or any spectral steepening
due to the high-energy protons that are no longer confined to the cluster \citep{1996SSRv...75..279V,1997ApJ...487..529B, 2010MNRAS.409..449P}, and thus, as discussed in the following, our results should be considered as  conservative. 
While this is not relevant when comparing with the \emph{Fermi} data, it might be relevant for the high-energy neutrino flux. 

Since the larger contribution to the total diffuse intensity comes from nearby galaxy clusters (see Figure~\ref{fig:results_1} and comments therein), we additionally omit the absorption of high-energy gamma rays owing to interactions with the extragalactic background light because this becomes relevant only at high redshifts (see, e.g., \citealp{2011MNRAS.410.2556D}). We note that our conclusions do not change even when relaxing any of the above approximations.

\section{Phenomenological luminosity-mass relation}
\label{sec:3}
In this section, we estimate the maximum possible contribution  to the extragalactic
gamma-ray and neutrino backgrounds from hadronic interactions in galaxy clusters using a simplified phenomenological approach
for the luminosity-mass relation. 

\subsection{Modelling the diffuse gamma-ray intensity}
\label{sec:3.1}
\begin{figure}[hbt!]
\centering
\includegraphics[width=0.5\textwidth]{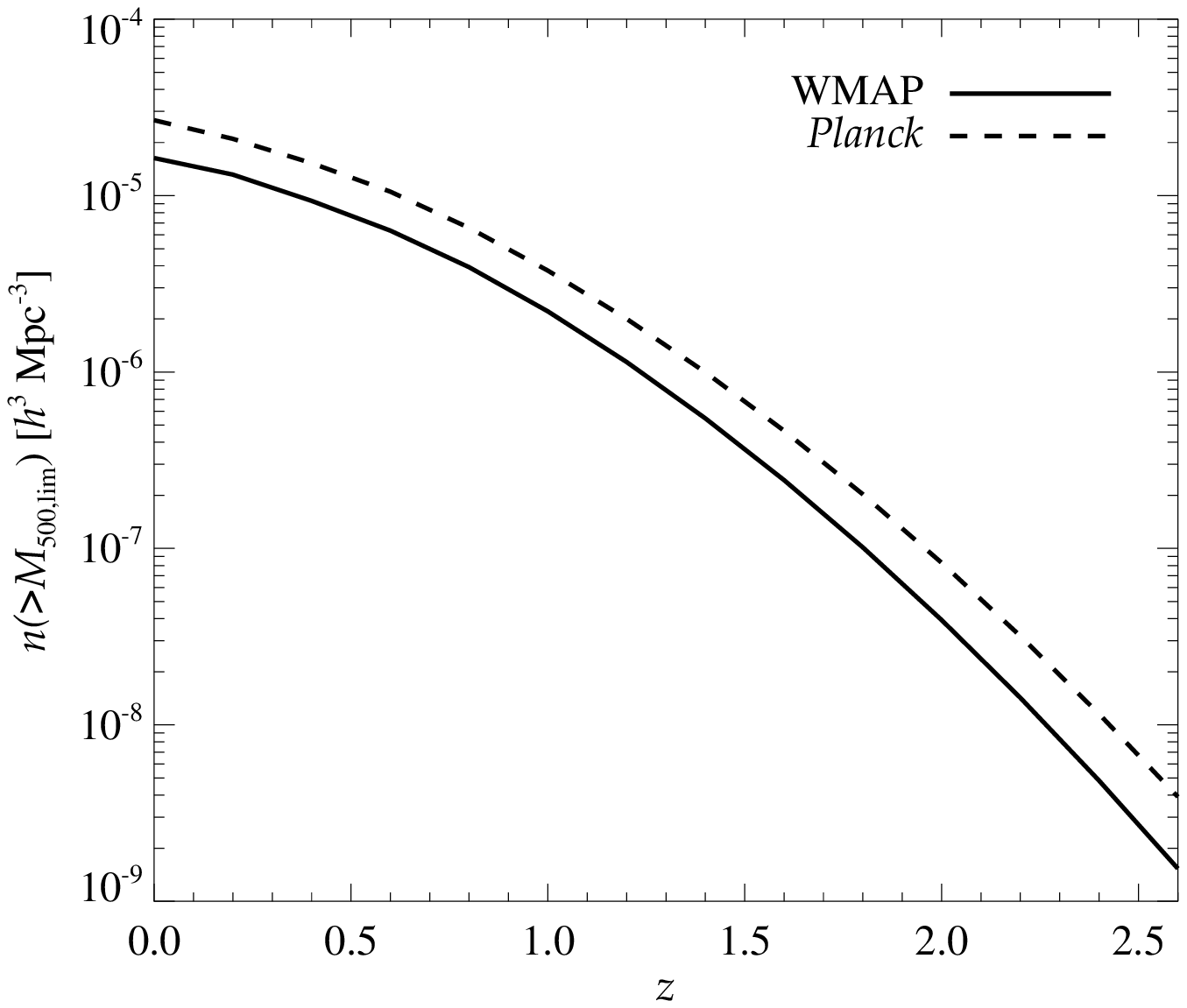}
\caption{Total number density of galaxy clusters for masses above
 $M_{500,\,\rmn{lim}}=10^{13.8}$~$h^{-1} M_\odot$ as a function of redshift. 
We show the number density obtained assuming the WMAP \citep{2011ApJS..192...18K}, our standard choice if
not otherwise specified, and the \emph{Planck} \citep{2013arXiv1303.5076P} cosmological data. 
At redshift $z = 2$, the number density is already negligible with respect to the lowest redshift.
}
\label{fig:massfun}
\end{figure}

The total gamma-ray intensity from all galaxy clusters in the Universe
at a given energy (d$N$\,/\,d$A$\,d$t$\,d$E$) is 
\begin{eqnarray}
I_{\gamma} = \int_{z_1}^{z_2} \int_{M_{500,\,\rmn{lim}}} & &
 \frac{L_{\gamma}(M_{500},z)\,(1+z)^2}{4\pi D_{\rm L}(z)^2} \\ \nonumber
& \times & \frac{\rmn{d}^{2}n (M_{500},z)}{\rmn{d}V_{\rmn{c}} \, \rmn{d}M_{500}} \frac{\rmn{d}V_{\rmn{c}}}{\rmn{d}z} \rmn{d}z \, \rmn{d}M_{500}  \, ,
\label{eq:TOTgamma}
\end{eqnarray}
where the cluster mass $M_{\Delta}$ is defined with respect to a density that is $\Delta=500$ times 
the \emph{\emph{critical}} density of the Universe at redshift $z$. Here,
$V_{\rmn{c}}$ is the comoving volume, $D_{\rm L}(z)$  the luminosity distance, and 
$\rmn{d}^{2}n (M_{500},z)/\rmn{d}V_{\rmn{c}} \, \rmn{d}M_{500}$ is the cluster mass function for which
we make use of the \cite{2008ApJ...688..709T} formalism and the \cite{2013A&C.....3...23M} on-line
application. The lower limit of the mass integration has been chosen to be $M_{500,\,{\rm lim}} = 10^{13.8}$~$h^{-1}$~M$_{\odot}$, to account for large galaxy groups. The redshift integration goes 
from $z_1 = 0.01$, where the closest galaxy clusters are located, up to $z_2 = 2$. Where not otherwise
specified, we assume
$\Omega_{\rmn{m}}=0.27$, $\Omega_{\Lambda}=0.73$, and the Hubble parameter
$H_0 = 100\,h_{70}$~km~s$^{-1}$~Mpc$^{-1}$ with $h_{70}=0.7$.
Where we explicitly use $h$ in the units, as for $M_{500,\,\rmn{lim}}$, we assume 
$H_0 = 100\,h$~km~s$^{-1}$~Mpc$^{-1}$ with $h=1$.
As shown in Figure~\ref{fig:massfun} (and discussed in Sections~\ref{sec:3.3} and \ref{sec:4.3}), 
our conclusions are not affected by the specific choice of $z_2$ and $M_{500,\,\rmn{lim}}$.

We calculate the total number of detectable galaxy clusters at $f = 1.4$~GHz, above the flux $F_{\rmn{min}}$, 
as
\begin{equation}
N_{1.4} (>F_{\rmn{min}}) = \int_{z_1}^{z_2} \int_{F_{\rmn{min}}}^{\infty} \frac{\rmn{d}^{2}n (F_{1.4},z)}{\rmn{d}V_{\rmn{c}} \, \rmn{d}F_{1.4}} \frac{\rmn{d}V_{\rmn{c}}}{\rmn{d}z} \rmn{d}z \, \rmn{d}F_{1.4}  \, ,
\label{eq:Nradio}
\end{equation}
 where $F_{1.4} = L_{1.4} (1+z) / 4 \pi D_{\rm L}(z)^2$, and we compare it with the radio counts from the National Radio Astronomy 
Observatory  Very Large Array sky survey (NVSS) of \cite{1999NewA....4..141G}.\footnote{We use the
cumulative number density function as in \cite{2010A&A...509A..68C}. \cite{2010A&A...509A..68C} 
do not use the fluxes of \cite{1999NewA....4..141G}, but rather the ones  from follow-up observations of the same sample of galaxy clusters, which are 
higher than the NVSS ones (R.~Cassano, private communication).} 
The flux $F_{\rmn{min}}$ is defined as in equation~(9) of  \cite{2012arXiv1210.1020C} by adopting a noise-level multiplier
$\xi_1 = 1$, which is appropriate, while slightly optimistic, for the low redshifts of the NVSS survey ($0.44\leq z \leq 0.2$), 
and a typical radio half-light radius of $R_{500}/4$ \citep{2012arXiv1207.6410Z}. 

The function $\rmn{d}^{2}n (F_{1.4},z)/\rmn{d}V_{\rmn{c}} \, \rmn{d}F_{1.4}$ is obtained numerically from 
$\rmn{d}^{2}n (M_{500},z)/\rmn{d}V_{\rmn{c}} \, \rmn{d}M_{500}$ by calculating $L_{1.4}(M_{500})$ from
$L_{\gamma}(M_{500})$ as explained in the following. 
We introduce a 
phenomenologically-driven gamma-ray luminosity-mass relation:
\begin{equation}
\rmn{log}_{10} \left[\frac{L_{\gamma} (100\,\rmn{MeV})}{\mathrm{s^{-1}\,GeV^{-1}}}\right] = P_{1} + P_{2} \, \rmn{log}_{10}
\left(\frac{M_{500}}{\mathrm{M}_\odot}\right) \, ,
\label{eq:LM}
\end{equation}
where we omit the possible redshift-dependence for simplicity.\footnote{Because the larger contribution to both the
number of detectable clusters in radio~\citep{2012arXiv1207.6410Z} and the total gamma-ray and neutrino fluxes
 is dominated by nearby clusters, the high-redshift dependence
is negligible for our purposes (see Sections~3.3 and 4.3 for more details). }
The radio luminosity can be obtained from the gamma-ray one by Equations~(\ref{eq:Lradio}) and (\ref{eq:Lgamma}).

In this section we assume that the magnetic field is independent of the
radius in the radio-emitting region. Therefore, the relation between radio and gamma-ray luminosities 
becomes
\begin{equation}
\frac{L_{\gamma}}{L_{f}} = \frac{A_{\gamma}}{A_{f}} \frac{\epsilon_{B}+\epsilon_{\rmn{CMB}}}{\epsilon_{B}} 
\left( \frac{\epsilon_{B_\rmn{c}}}{\epsilon_{B}}\right)^{\frac{\alpha_{\rmn{p}-2}}{4}}\, .
\label{eq:LR2}
\end{equation}
A special limit can be obtained for $B \gg B_{\rmn{CMB}}$ in all the radio-emitting region. In this case, under 
the hypothesis that electrons lose all their energy through synchrotron emission and $\alpha_\rmn{p} \approx 2$, 
the relation between radio and gamma-ray luminosities becomes \citep{2008MNRAS.385.1242P}:
\begin{equation}
\frac{L_{\gamma}}{L_{f}} \approx \frac{A_{\gamma}}{A_{f}} \, .
\label{eq:LR1}
\end{equation}

Concerning  the choice of the parameters in Equation~(\ref{eq:LM}), we
need to consider that $P_1$, $P_2$, $\alpha_{\rm p}$, $B$, and the fraction of 
loud clusters are degenerate when one tries to find the maximum allowed
hadronic-induced emission. 
The concept of loud fraction comes from the fact that, even if clusters
have the same X-ray luminosity and therefore the same mass, some of
them host radio emission, but others do not show any sign of it with
upper limits about an order of magnitude below the loud state.
This is known as the radio--X-ray bimodality 
\citep{2009A&A...507..661B,2013arXiv1306.4379C}. The most recent estimates 
suggest that the radio-loud percentage is about 20--30\%~\citep{2013arXiv1306.3102K}. 
The subdivision of the cluster population into radio-loud and radio-quiet clusters
is also reflected in the corresponding gamma-ray and neutrino fluxes. Therefore,  
from now on we refer to the two populations as ``loud'' and ``quiet.''

In this section we mainly consider the overly optimistic case where all the clusters
are loud (100\% loud), while we show the case of 30\% loud clusters for only one choice 
of $\alpha_{\rm p}$. In the following, to reduce the number of free parameters, we
fix $P_2 = 5/3 \simeq 1.67$; i.e., we assume that the hadronic-induced luminosity scales as the 
cluster thermal energy $E_{\rm th} \propto M^2/R_{\rm vir} \propto M^{5/3}$ (see also Section~3.3), where $R_{\rm vir}$ is the virial radius. 
The chosen $P_2$ parameter roughly corresponds to what is found using the  \cite{2013arXiv1311.4793Z} multi-frequency 
mock cluster catalogue (MultiDark database; \citealp{2011arXiv1109.0003R}) for $L_{\gamma} (100\,\rmn{MeV})$--$M_{500}$, which typically lies 
in the range $\sim$1.5--1.65 for different redshifts and different cluster populations (loud, quiet, cool-core, 
non-cool-core). The parameter $P_1$ is set free to vary under the constraint that it should respect the radio 
counts from the NVSS survey and current gamma-ray upper limits. We note that, once the thermal content 
of a cluster is known, the parameter $P_1$ could be seen as the efficiency of how much energy goes into CR acceleration.

We considered the Coma and Perseus cases for comparison with current gamma-ray upper limits on individual galaxy clusters.
We took the Coma upper limit obtained from five years of \emph{Fermi} data by \cite{2014MNRAS.440..663Z} as reference. We adopted
their result for the disk model, a uniform filling of the cluster up to $R_{200}$, which is $F_{\rmn{UL}}(>100~\rmn{MeV}) = 2.9\times10^{-9}$~cm$^{-2}$~s$^{-1}$,
obtained for a spectral index of $2$. For Perseus, we assumed the upper limit obtained by the \cite{2011arXiv1111.5544M} for the inner region of $0\fdg15$ as reference,
which is $F_{\rmn{UL}}(>1~\rmn{TeV}) = 1.4\times10^{-13}$~cm$^{-2}$~s$^{-1}$, obtained for a spectral index of $2.2$.
We refer the reader to, for example, Table~1 of \cite{2013arXiv1308.6278H} and Table~2 of \cite{2010ApJ...710..634A} for hints to how much the 
gamma-ray upper limits change when modifying the spectral index. Such a change is quantifiable within a factor of about two, which
does not affect our conclusions, as we discuss later.

\begin{table*}[t]
\begin{center}
  \caption{\label{tab:LM_trials} Tested parameters and total gamma-ray and neutrino fluxes for the phenomenological luminosity-mass relation.
    }
\resizebox{\textwidth}{!}{
\begin{tabular}{ccccccccc}
\hline\hline
\phantom{\Big|}
$\alpha_{\rmn{p}}$ & Loud [\%] & $B$ [$\mu$G] & $P_{1}$ & Coma ($>100$~MeV) & Perseus ($>1$~TeV) & $I_{\gamma}$ (100~MeV) & $I_{\nu}$ (250~TeV) &Notes\\
\hline\\[-0.5em]
1.5 & 100 &  $\gg B_{\rmn{CMB}}$  &  18.60 (18.35)  & 1.6~(0.92)~$\times10^{-11}$  & 1.7~(0.92)~$\times10^{-13}$  & 3.8~(2.1)~$\times10^{-10}$ &  7.3~(4.2)~$\times10^{-19}$ & N \\
      &         &  1                                  &  19.41 (18.35)  & 1.1~(0.09)~$\times10^{-10}$   & 1.1~(0.09)~$\times10^{-12}$  & 2.5~(0.2)~$\times10^{-9}$ & 4.7~(0.4)~$\times10^{-18}$ & N\\
      &         &  0.5                               &  19.91 (18.35)  & 3.3~(0.09)~$\times10^{-10}$   & 3.4~(0.09)~$\times10^{-12}$  & 7.8~(0.2)~$\times10^{-9}$ & 1.5~(0.04)~$\times10^{-17}$ & N\\
\hline\\[-0.5em]
2    & 100 &  $\gg B_{\rmn{CMB}}$  &  19.42  & $6.0\times10^{-11}$  & $1.1\times10^{-14}$  & $2.5\times10^{-9}$ & $4.7\times10^{-21}$ & \\
      &        &  1                                   &  20.65  & $1.0\times10^{-9}$   & $1.8\times10^{-13}$  & $4.3\times10^{-8}$ & $8.1\times10^{-20}$ & \\
      &        &  0.5                                &  21.23 (21.09)  & 3.9~(2.8)~$\times10^{-9}$  & 6.9~(5.0)~$\times10^{-13}$  & 1.6~(1.2)~$\times10^{-7}$ & 3.1~(2.2)~$\times10^{-19}$ & G\\
2    &  30  &  $\gg B_{\rmn{CMB}}$  &  19.60  & $9.1\times10^{-11}$  & $1.6\times10^{-14}$  & $1.4\times10^{-9}$ & $2.8\times10^{-21}$ & \\
      &        &  1                                   &  20.82  & $1.5\times10^{-9}$   & $2.7\times10^{-13}$  & $2.3\times10^{-8}$ & $4.6\times10^{-20}$ & \\
      &        &  0.5                                &  21.40 (21.09)  & 5.7~(2.8)~$\times\,10^{-9}$   & 1.0~(0.5)~$\times\,10^{-12}$  & 8.9~(4.4)~$\times\,10^{-8}$ & 1.8~(0.9)~$\times\,10^{-19}$ & G\\
\hline\\[-0.5em]
2.2 & 100 &  $\gg B_{\rmn{CMB}}$  &  19.71  & $1.0\times10^{-10}$  & $3.6\times10^{-15}$  & $4.9\times10^{-9}$ & $5.9\times10^{-22}$ & \\
      &        &  1                                   &  21.10  & $2.6\times10^{-9}$   & $8.7\times10^{-14}$  & $1.2\times10^{-7}$ & $1.4\times10^{-20}$ & \\
      &        &  0.5                                &  21.71 (21.16)  & 1.0~(0.3)~$\times\,10^{-8}$   & 3.6~(1.0)~$\times\,10^{-13}$  & 4.9~(1.4)~$\times\,10^{-7}$ & 5.9~(1.7)~$\times\,10^{-20}$ & G\\
\hline\\[-0.5em]
2.4 & 100 &  $\gg B_{\rmn{CMB}}$  &  19.98  & $1.6\times10^{-10}$  & $1.2\times10^{-15}$  & $9.1\times10^{-9}$ & $7.0\times10^{-23}$ & \\
      &        &  1                                   &  21.54 (21.21)  & 5.9~(2.8)~$\times10^{-9}$ & 4.2~(1.9)~$\times10^{-14}$  & 3.3~(1.6)~$\times10^{-7}$ & 2.5~(1.2)~$\times10^{-21}$ & G\\
      &        &  0.5                                &  22.18 (21.21)  & 2.6~(0.3)~$\times\,10^{-8}$   & 1.8~(0.2)~$\times\,10^{-13}$  & 1.4~(0.2)~$\times\,10^{-6}$ & 1.1~(0.1)~$\times10^{-20}$ & G\\
\hline
\end{tabular}
}
\end{center}
\small{{\bf Note.} For each $\alpha_{\rmn{p}}$ and magnetic field, the $P_1$ parameter of the $L_{\gamma} (100\,\rmn{MeV})$--$M_{500}$,
obtained by taking the NVSS radio counts into account, is reported in the fourth column. Cols. 5 \&\ 6:\ corresponding 
Coma-like and Perseus-like gamma-ray flux in cm$^{-2}$~s$^{-1}$, respectively, integrated above 100~MeV and 1~TeV, and assuming the clusters 
$M_{500}$ as in \cite{2002ApJ...567..716R}. Cols 7 \& 8:\ total gamma-ray and neutrino (all flavours) intensity 
at 100~MeV and 250~TeV, respectively, in cm$^{-2}$~s$^{-1}$~GeV$^{-1}$~sr$^{-1}$ for all the galaxy clusters in the Universe. Last 
column: ``G'' and ``N'', cases overshooting present gamma-ray and neutrino constraints, respectively. 
For $\alpha_{\rmn{p}} \ge 2$, we report in parenthesis the values that respect the gamma-ray upper limit on Coma, while for $\alpha_{\rmn{p}} = 1.5$ 
we report in parenthesis the values matching the IceCube neutrino data averaging in the corresponding energy range.} 
\end{table*}

\subsection{Results: gamma-ray and neutrino backgrounds}
\label{sec:3.2}
We assume the spectral index $\alpha_{\rmn{p}} = 2$, 2.2, 2.4 and, as extreme case, $\alpha_{\rmn{p}} = 1.5$.
As for the magnetic field $B \gg B_{\rmn{CMB}}$ (see Equation~\ref{eq:LR1}), $B = 1$~$\mu$G, and $0.5$~$\mu$G 
(see Equation~\ref{eq:LR2}). The first choice of the magnetic field can be regarded as conservative considering
that, for example, the volume-averaged magnetic field of Coma, the best-studied cluster for Faraday rotation 
measurements, is about 2~$\mu$G \citep{2010A&A...513A..30B}; the latter should be considered 
optimistic with respect to current estimates. To clarify the meaning of the terms conservative/optimistic, note 
that the higher the magnetic field, the less room there is for protons, because the radio counts have to be respected, 
hence the lower the gamma-ray and neutrino fluxes.

For each $\alpha_{\rmn{p}}$ and value of the magnetic field, the corresponding $P_1$ parameter
is chosen in such a way that the computed $N_{1.4}(>F_{\rm min})$ does
not overshoot the NVSS radio counts, and they are reported in
Table~\ref{tab:LM_trials}. 
To make certain that our models respect current gamma-ray upper limits, the corresponding 
Coma-like and Perseus-like gamma-ray fluxes above 100~MeV and 1~TeV, respectively, are also shown in Table~\ref{tab:LM_trials}, 
after assuming $M_{500}$ as in \cite{2002ApJ...567..716R}, together with the total gamma-ray and neutrino flux at 
100~MeV and 250~TeV, respectively, for all the galaxy clusters in the Universe.  All the reported values refer to 100\% 
loud clusters, while the 30\% case is studied only for $\alpha_{\rmn{p}} = 2.$ (In the latter case, the remaining fraction of $70\%$ 
quiet clusters are assumed to have an $L_{\gamma} (100\,\rmn{MeV})$ that is one order of magnitude lower than the loud ones.) 

In the last column of Table~\ref{tab:LM_trials} and for $\alpha_{\rmn{p}} \ge 2$, we denote the cases that do not respect the gamma-ray upper limits on either Coma or Perseus by ``G''. 
For these cases, we recomputed $P_1$ so as to respect the \emph{Coma} upper limit, 
our reference choice (see values in parenthesis in Table~\ref{tab:LM_trials}). 
However,  our recomputed values  for $\alpha_{\rmn{p}} = 2$ still overshoot the current Perseus gamma-ray upper limit. 
We nevertheless adopt the Coma upper limit as reference because it was calculated for $\alpha_{\rmn{p}} = 2$ and for a larger 
spatial extension, up to $R_{200}$.
For $\alpha_{\rmn{p}} = 1.5$, the cases indicated by ``N'' in Table~\ref{tab:LM_trials} exceed the IceCube neutrino data. Also in this
case we recalculated $P_1$ to match the IceCube results after averaging over the corresponding energy range.

\begin{figure*}[hbt!]
\centering
\includegraphics[width=0.457\textwidth]{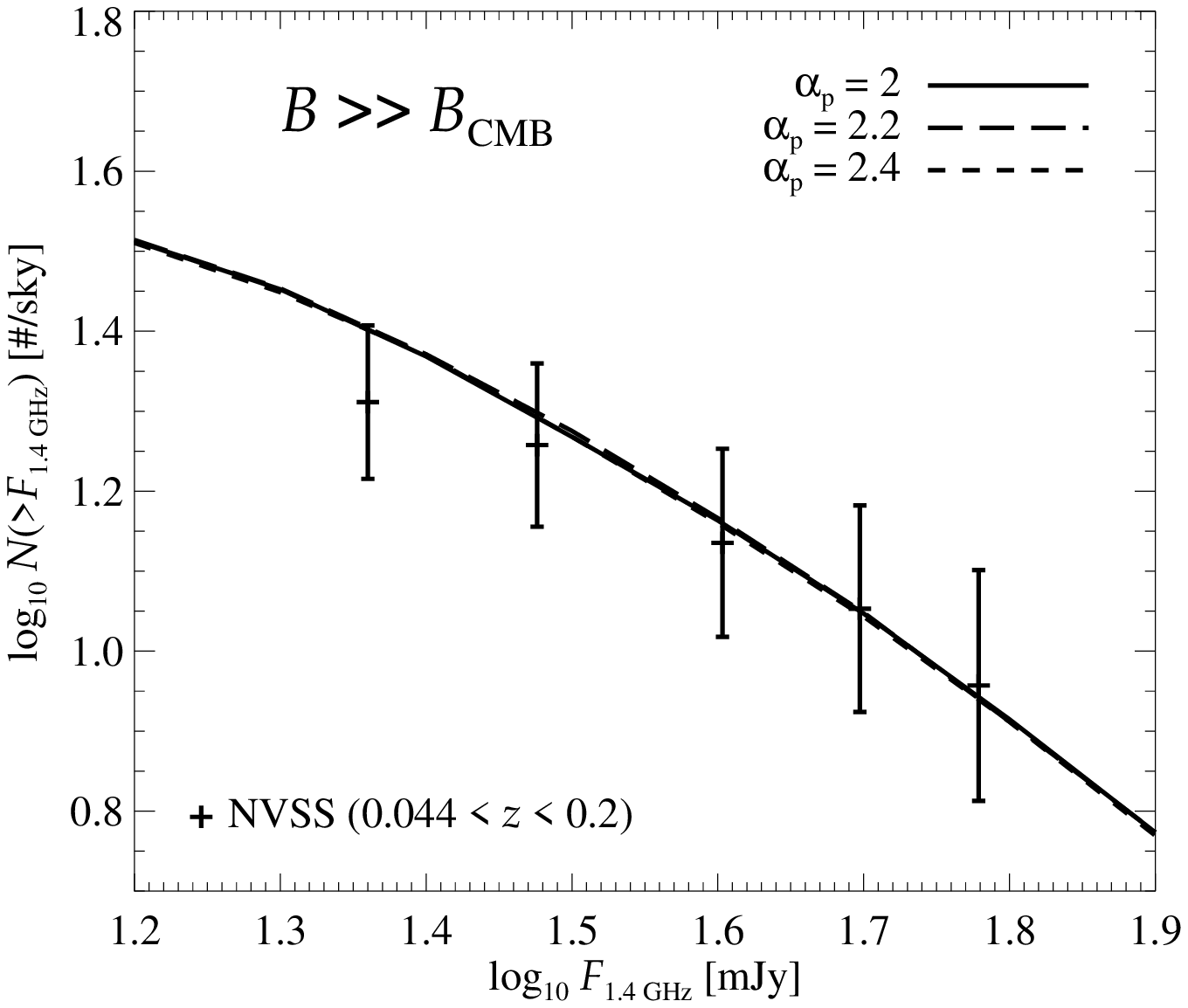}
\includegraphics[width=0.536\textwidth]{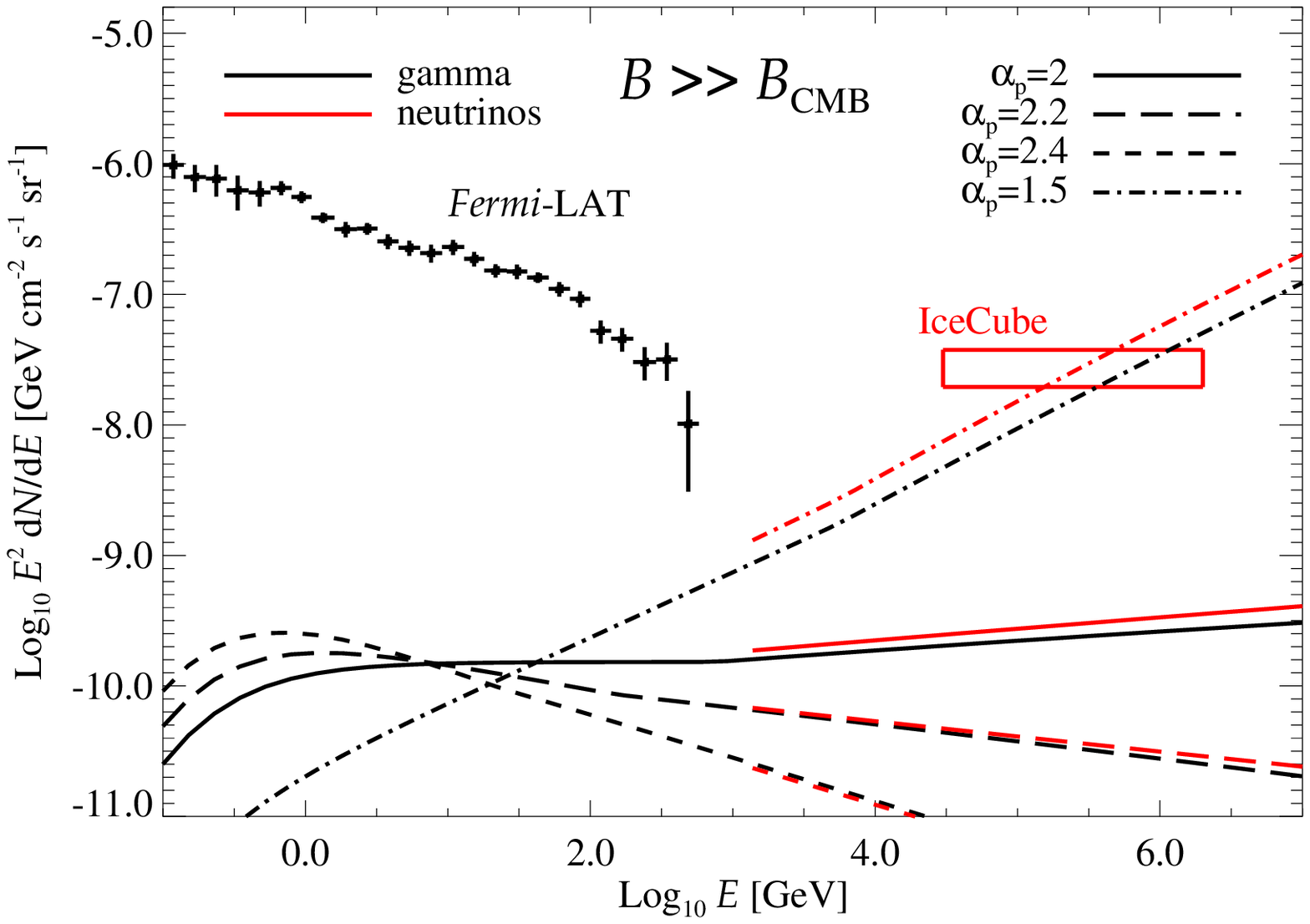}
\includegraphics[width=0.457\textwidth]{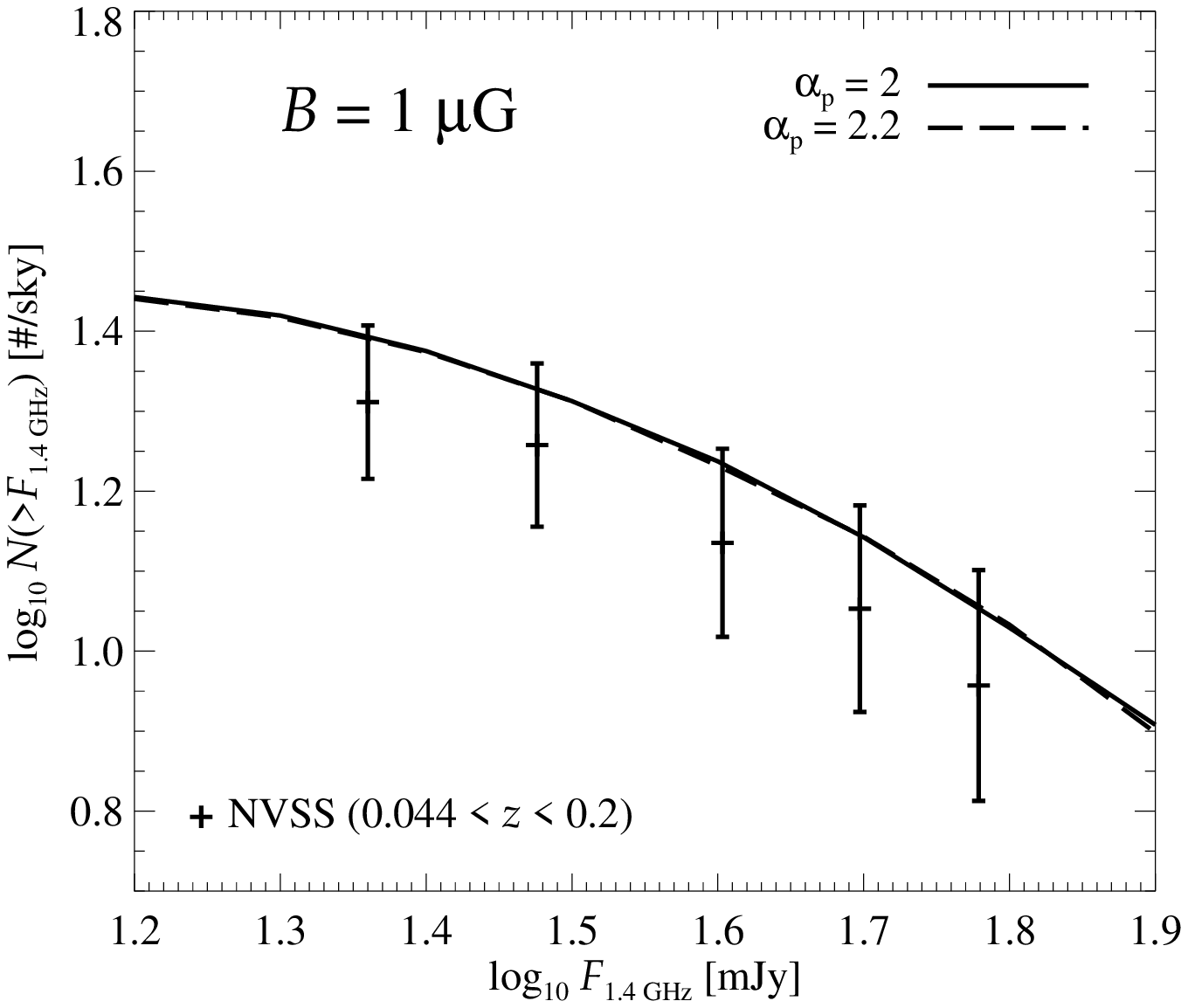}
\includegraphics[width=0.536\textwidth]{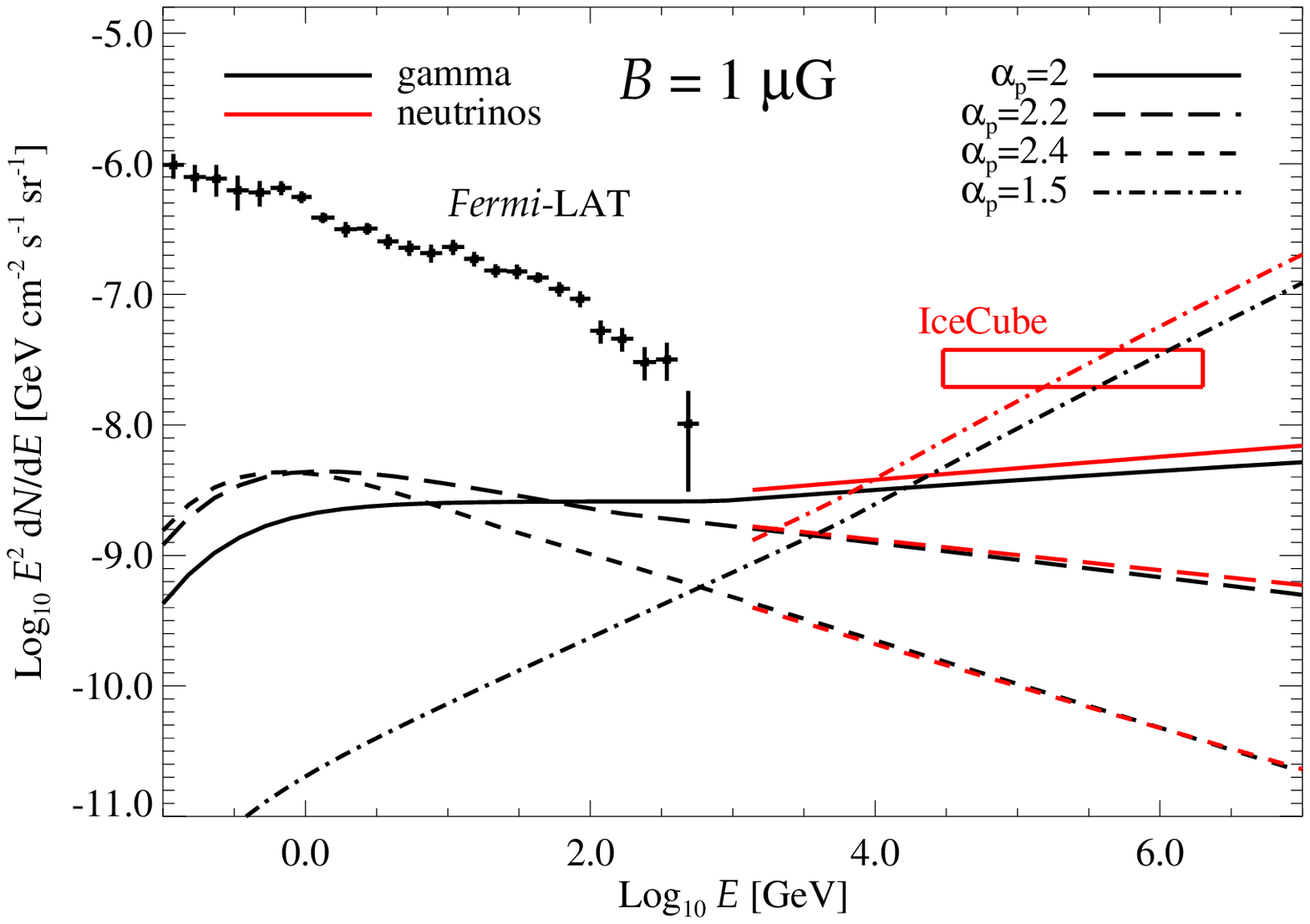}
\includegraphics[width=0.457\textwidth]{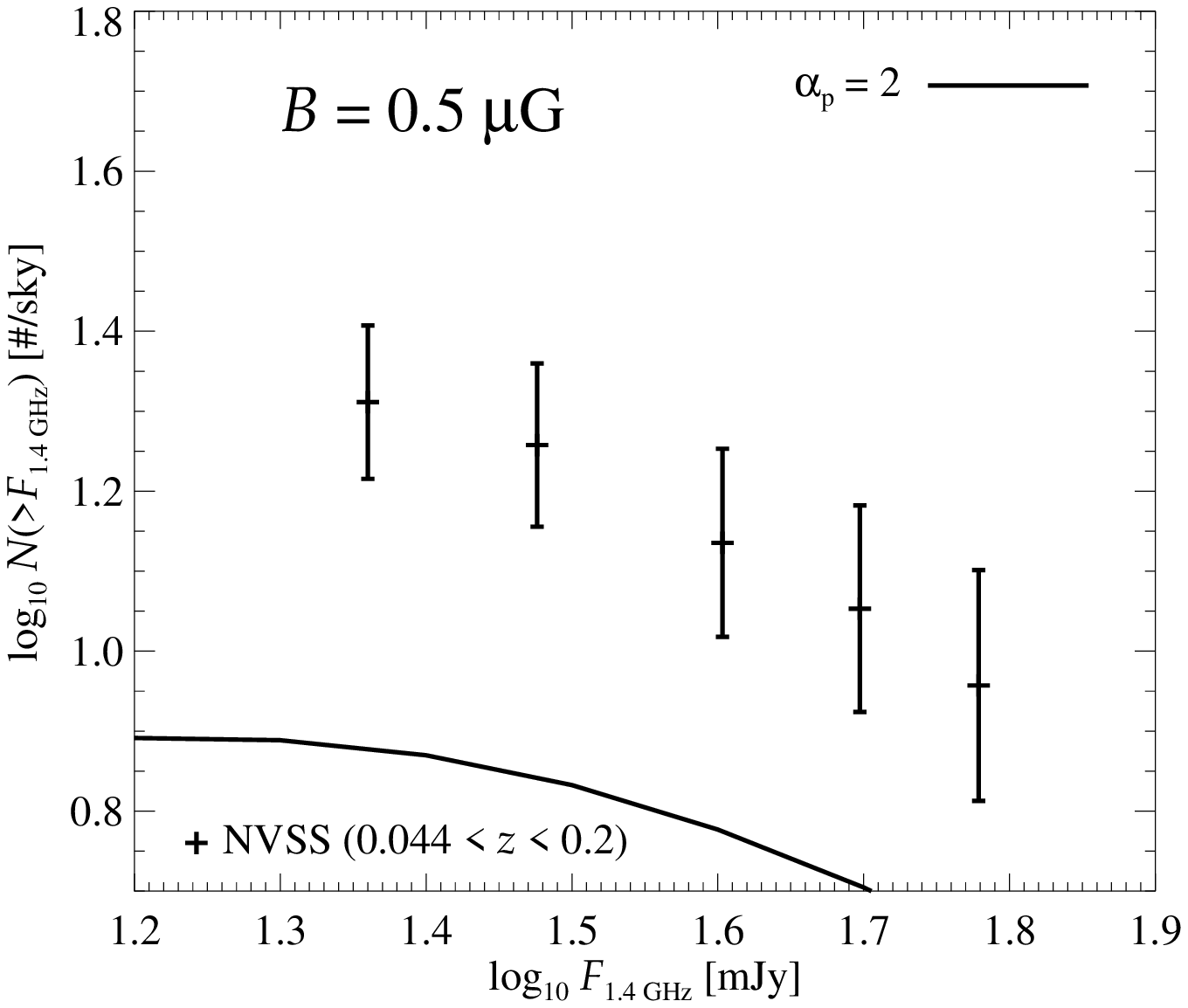}
\includegraphics[width=0.536\textwidth]{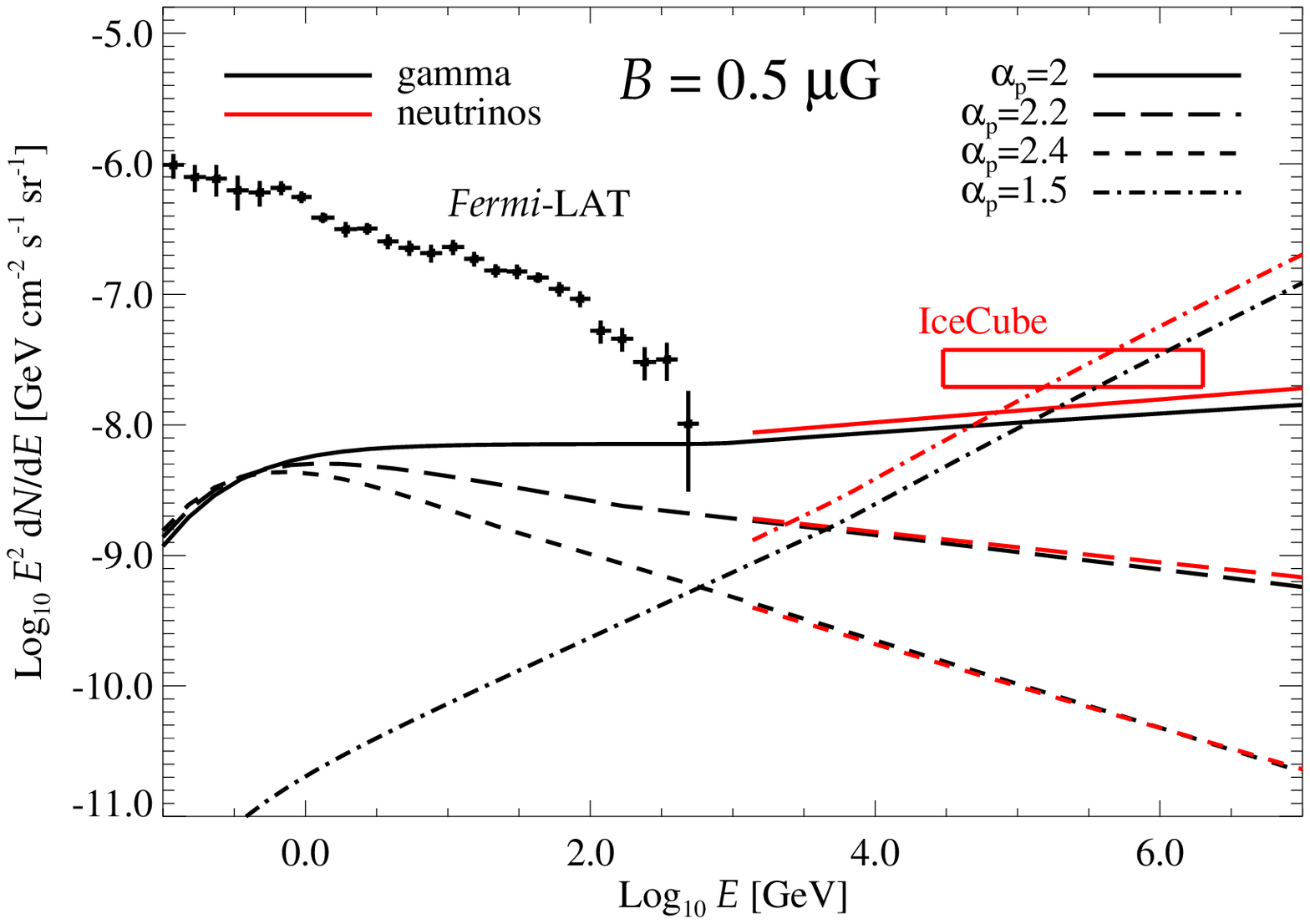}
\caption{
Total gamma-ray and neutrino intensities (right) due to hadronic interactions in galaxy clusters, for 100\% loud clusters, and 
the corresponding radio counts due to synchrotron emission from secondary electrons (left). From top to bottom, we plot the cases with
$B\gg B_{\rmn{CMB}}$, $B = 1$~$\mu$G and $0.5$~$\mu$G, respectively. For comparison,  the {\it Fermi}~\citep{2014arXiv1410.3696T} 
and IceCube~\citep{Aartsen:2014gkd} data are shown in the panels on the right. 
The neutrino intensity is meant for all flavours. All the plotted intensities respect NVSS radio counts and the 
gamma-ray upper limits on individual clusters. For $B = 1$~$\mu$G and $\alpha_{\rmn{p}} = 2.4$,
$B = 0.5$~$\mu$G and $\alpha_{\rmn{p}} = 2.2, 2.4$, and for $\alpha_{\rmn{p}} = 1.5$, the radio counts respecting 
the gamma-ray and neutrino limits, respectively, are below the y-scale range adopted for the panels on the left.
}
\label{fig:resultsLM1}
\end{figure*}

Figure~\ref{fig:resultsLM1} shows both the comparison of our models to the radio counts (on the left) and the computed 
gamma-ray (in black) and neutrino intensities (in red) as functions of the energy (on the right), for the chosen values of 
$\alpha_{\rmn{p}}$ and $B$ assuming 100\% loud clusters. For comparison, we plot the {\it Fermi} data \citep{2014arXiv1410.3696T}
and the IceCube 1$\sigma$ error band as in~\cite{Aartsen:2014gkd}. The latter refers to the four-year IceCube
data sample. However, more recently a new fit has been provided, using  two-year statistics but including low energy 
events down to $1$~TeV. The best fit of the neutrino spectrum obtained in this case scales as $E_{\nu}^{-2.46}$ \citep{Aartsen:2014yta}.

For $\alpha_\rmn{p} > 2$, both the gamma-ray and the neutrino diffuse backgrounds are well below the {\it Fermi} 
and the IceCube data in all cases. For $\alpha_\rmn{p} = 2$, while the gamma-ray flux is always lower 
than the {\it Fermi} measurements, the neutrino diffuse background could represent a significant fraction of
the flux measured by IceCube for $B = 1$~$\mu$G and 0.5~$\mu$G. 

\begin{figure*}[hbt!]
\centering
\includegraphics[width=0.457\textwidth]{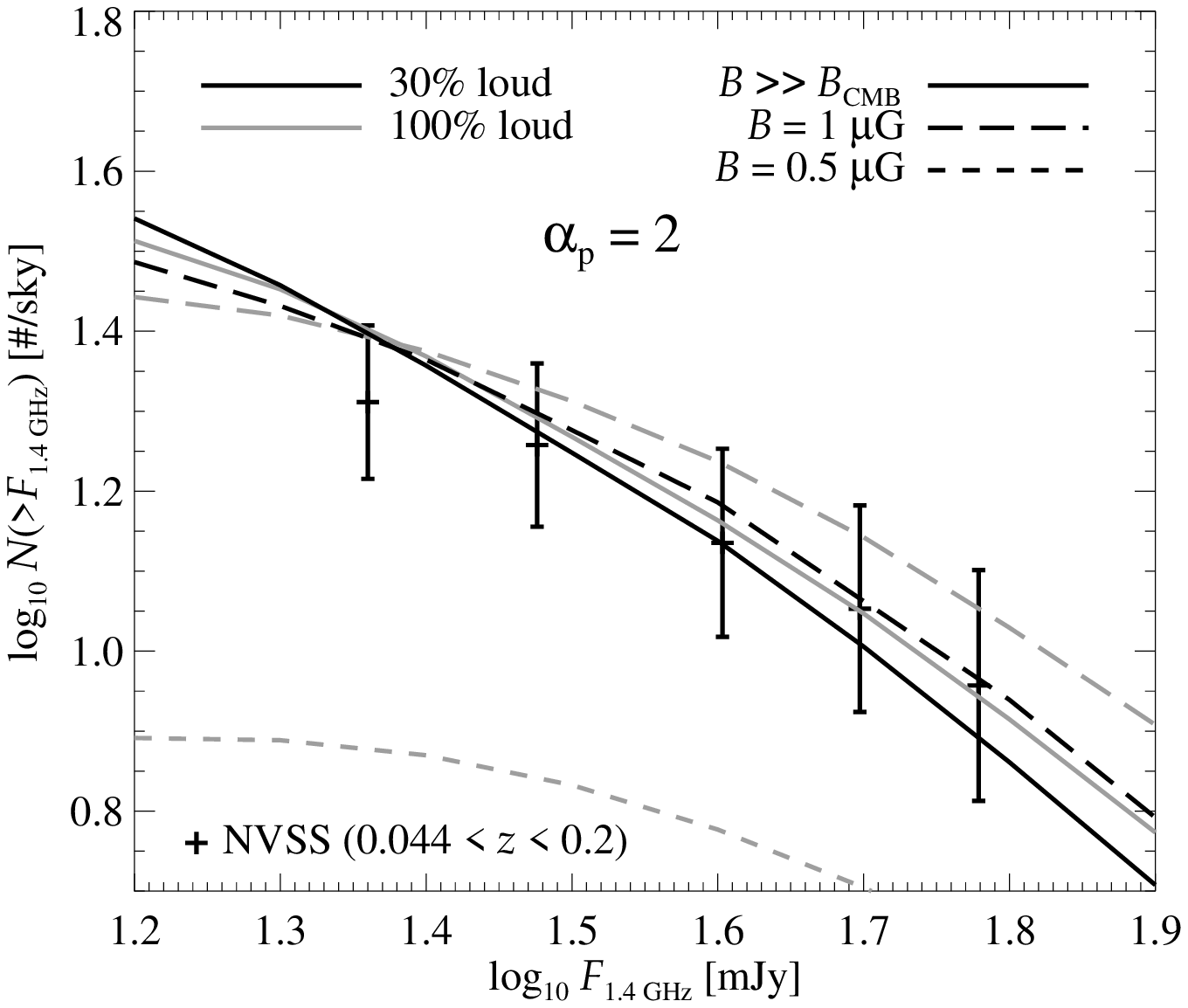}
\includegraphics[width=0.536\textwidth]{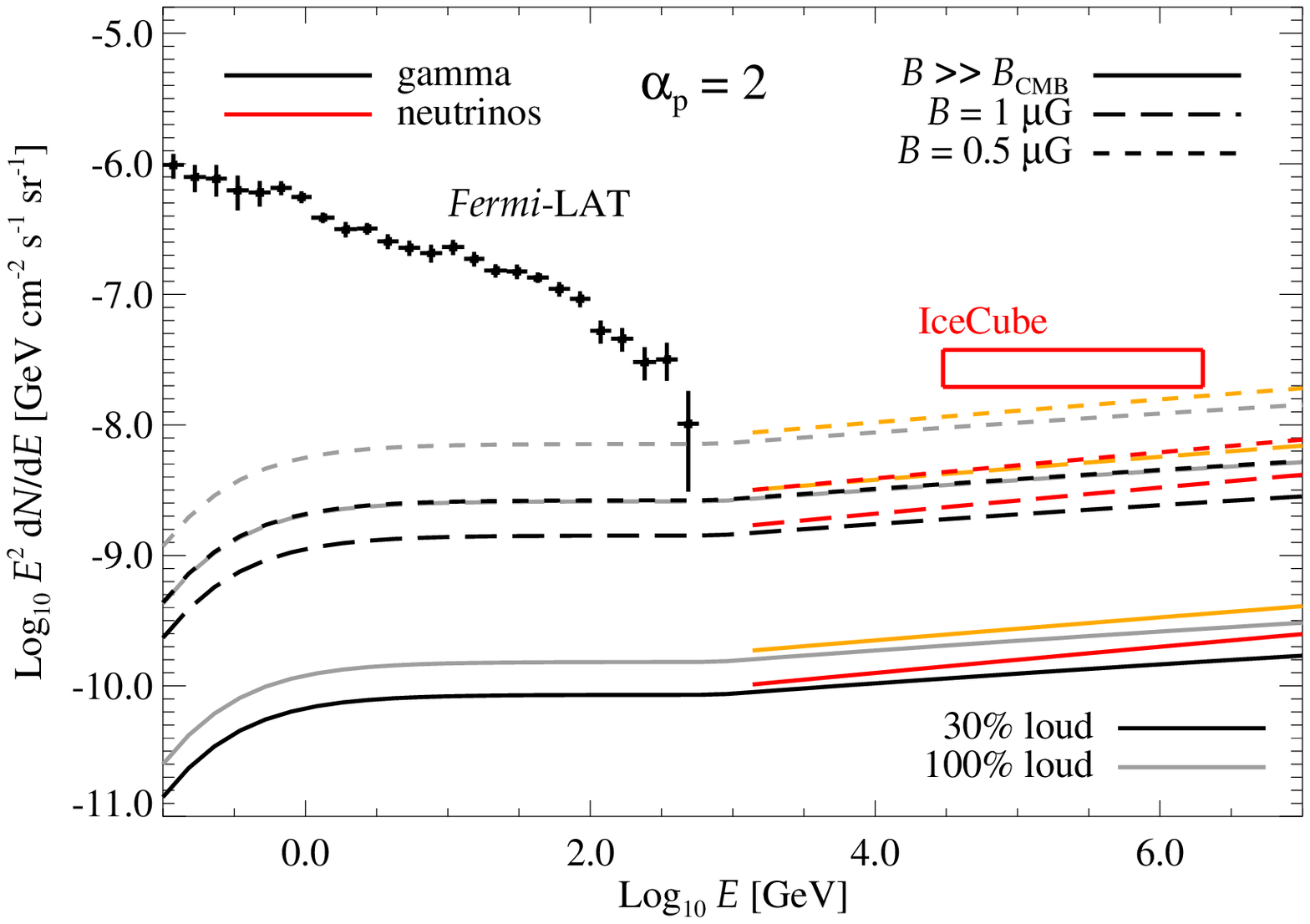}
\caption{
Same as Figure~\ref{fig:resultsLM2}, together with the case of 30\% loud clusters for $\alpha_{\rmn{p}} = 2$. 
The remaining percentage of $70\%$ quiet clusters has been assumed to have $L_{\gamma} (100\,\rmn{MeV})$ one order of 
magnitude lower than for the loud clusters. The 100\% loud case is shown with lighter colours (i.e., in grey and orange).
}
\label{fig:resultsLM2}
\end{figure*}

As known from radio observations, the case of 100\% loud clusters is not realistic. Therefore, in Figure~\ref{fig:resultsLM2}, 
we show the same as in Figure~\ref{fig:resultsLM1} for $\alpha_\rmn{p} = 2,$ together with the more realistic 
case of 30\% loud clusters. In the latter, galaxy clusters could make up at most about 10\% (20\%)
of the neutrino flux measured by Ice Cube for $B = 1$~$\mu$G (0.5~$\mu$G). This gives an estimation of how much our results for 
other spectral indices would change when moving from 100\% loud clusters to the more realistic 
case of 30\% loud clusters: $I_{\gamma,\nu,30\%} \approx I_{\gamma,\nu,100\%}/2$ (see also Table~\ref{tab:LM_trials} for comparison).

In the extreme case of $\alpha_\rmn{p} = 1.5$, we could explain the IceCube data by averaging over the corresponding energies
for all cases, while respecting all other constraints from radio to gamma rays. However, we note that such a hard spectral index contradicts
the most recent IceCube results, thus suggesting a softer spectral index \citep{Aartsen:2014yta}.

Estimates of magnetic fields in clusters from Faraday rotation measurements range from $\sim \mu$G for merging clusters 
up to 10~$\mu$G for cool-core clusters \citep{2002ARA&A..40..319C,2004JKAS...37..337C,2005A&A...434...67V,
2010A&A...513A..30B,2013MNRAS.tmp.1637B}. The case of $B = 0.5$~$\mu$G should therefore be considered 
illustrative and optimistic because it contradicts current knowledge.

We conclude that, amongst all the cases we studied 
that respect both radio counts and current gamma-ray upper limits, hadronic interactions in galaxy clusters can 
realistically contribute at most up to $10$\% of the total extragalactic neutrino background, while contributing 
less than a few percentage points to the total extragalactic gamma-ray background. Moreover, the simplified requirement of not
overshooting the NVSS radio counts on clusters leads to optimistic results. In fact, as explained in 
Section~\ref{sec:2}, not all the observed radio emission in clusters has a hadronic origin 
\citep{2012arXiv1207.3025B,2012arXiv1207.6410Z}. The open question 
is the exact contribution of protons to the non-thermal content of clusters, the corresponding contribution 
to the observed radio emission, and therefore, the possible gamma-ray emission (see \citealp{2014MNRAS.440..663Z} for a discussion). 
This implies that even our results, which respect both NVSS counts and gamma-ray limits, should still be considered rather optimistic.

Finally, we note that, owing to our simplified approach using a gamma-ray luminosity--mass relation, 
the conclusions of this section can be generalised to any source of CR protons where these mix 
and hadronically interact with the ICM of galaxy clusters, such as those injected by structure formation 
shocks and AGNs. For any considered source of protons, the resulting secondary emission must 
respect both radio and gamma-ray constraints.

\subsection{Results: testing our standard assumptions}
\label{sec:3.3}
To make our conclusions more robust, we comment in this section on two of our assumptions and on their effect on our final results: the redshift 
evolution and the value of the parameter $P_2$ in the luminosity-mass relation.

The redshift dependence has been omitted in Equation~\ref{eq:LM}. For the sake of completeness, we tested the effect of introducing a redshift
dependence in the gamma-ray luminosity-mass relation by adopting $L_\gamma \propto \Omega_{\rmn{m}}\,(1+z)^{3} + \Omega_{\Lambda}$, 
for  $\alpha_{p} = 2.2$, 100\% loud clusters and $B \gg B_{\rmn{CMB}}$, roughly corresponding to the scaling observed in 
the \cite{2013arXiv1311.4793Z} multi-frequency mock-cluster catalogue for $L_{\gamma} (100\,\rmn{MeV})$--$M_{500}$. 
We found that omitting the redshift evolution causes both the radio counts and the high-energy fluxes to be only about 20\% 
lower than the redshift-evolution case. Our results would scale accordingly, as would the $P_1$ parameter, and 
the maximum allowed contribution to the total extragalactic gamma-ray and neutrino fluxes would remain approximately 
the same. The case of radio counts is not as intuitive, but can be understood if noting that redshift evolution will boost the 
luminosity of higher redshift objects, pushing them into a regime where they would be detectable and boosting the corresponding 
estimation of the radio counts, therefore requiring lower $P_1$ with respect to no-redshift evolution.

In Section~\ref{sec:3.1}, we fix the slope of the luminosity-mass relation to $P_2 = 5/3$, assuming that the 
hadronic-induced luminosity scales as the cluster thermal energy. In the conclusions of Section~\ref{sec:3.2},
we mentioned that our phenomenological approach can be generalised to any source of CR protons in
clusters if these mix and hadronically interact with the ICM. However, while our standard choice for the $P_2$ parameter 
is appropriate for CR protons injected by structure formation shocks, it could be different for other CR sources. Clearly, 
a steeper slope would assign larger fluxes to high-mass objects that would easily overshoot radio counts. As a 
consequence, a lower value for $P_1$ would be allowed, and considering that low-mass clusters would have lower
luminosities, we estimate that the total gamma-ray and neutrino fluxes would be lower than in the case of
$P_2 = 5/3$, or at most, at the same level owing to the sum of a few very powerful, massive nearby sources.

To assess the changes obtained by assuming a flatter slope in the luminosity-mass relation, we tested the
extreme value $P_2 = 1$ for 30\% loud clusters, $\alpha_\rmn{p} = 2$ and $B = 1$~$\mu$G, 
our most optimistic, still realistic, case. We underline, however, that a luminosity-mass function with such a flat 
slope strongly contradicts current knowledge of the diffuse radio emission in galaxy clusters~\citep{2009A&A...507..661B,2013arXiv1306.4379C}.
Either way, we found that, to respect radio counts, the maximum allowed contribution 
to the total extragalactic neutrino flux is about 15\%. This behaviour can be understood when noting again that such 
a flat slope implies that higher luminosities are assigned to lower mass clusters, pushing them into a regime where 
they would be detectable, hence boosting the corresponding radio counts. For the sake of completeness, we also 
added a redshift evolution of the luminosity as $(1+z)^3$ (as, e.g., for AGNs; \citealp{2005AJ....129..578B}) to 
this extreme model that should eventually boost the neutrino production. We found that the maximum allowed contribution to the total extragalactic 
neutrino flux is  30\% of the IceCube flux. We conclude that in all cases, the contribution to the total extragalactic gamma-ray
flux is still negligible.

The estimation of a flux that is 30\% of the IceCube one is the maximum that can be obtained under 
realistic conditions (30\% loud clusters, $B = 1$~$\mu$G) for the extreme value $P_2 = 1$ with $\alpha_\rmn{p}=2$. 
The only way to additionally boost the total neutrino flux without changing the radio counts would be to integrate down to lower 
masses, as we also discuss in Section~\ref{sec:4.3}.\footnote{In Section~\ref{sec:4.3} we also estimate the
changes obtained by adopting the most recent \emph{Planck} results for the cosmological parameters \citep{2013arXiv1303.5076P}.
While for the semi-analytical model of the next section, the radio counts, total gamma-ray, and neutrino 
fluxes are enhanced  by a factor of only about 1.7, in the phenomenological model with $P_2 = 1$, this 
would significantly boost the radio counts requiring the corresponding $P_1$ value to be lowered.} 
We note, however, that our standard lower mass bound is $M_{500,\,\rmn{lim}} = 6.3\times10^{13} h^{-1} M_\odot = 
9 \times10^{13} M_\odot$, roughly corresponding to $M_{200,\,\rmn{lim}} = 1.4 \times10^{14} M_\odot$, and it already includes 
groups of galaxies. Extending the mass integration of the above case down to $M_{500,\,\rmn{lim}} = 10^{13} h^{-1} M_\odot = 
1.4 \times10^{13} M_\odot$, the 30\% contribution to the total neutrino flux would become about 160\%, overshooting the
IceCube measurement. One could, of course, fine-tune this mass limit to match the IceCube flux, but we think that such a 
combination of extreme parameters is highly unlikely. At any rate, the $E^{-2}$ spectrum is 
the only one for which such fine-tuning would give a significant total neutrino flux, and it disagrees with the latest 
IceCube results, suggesting a softer spectral index \citep{Aartsen:2014yta}.

We conclude that the results of the phenomenological approach presented in Section~\ref{sec:3.2} are robust
against our assumptions and that they provide realistic estimates of the maximum allowed contribution of galaxy clusters to the total
extragalactic gamma-ray and neutrino fluxes.

\section{Semi-analytical model for the cosmic-ray and intra-cluster-medium distributions}
\label{sec:4}
In this section, we adopt a more sophisticated approach to modelling the CR and ICM distributions in galaxy clusters, as well as 
their magnetic field spatial distribution. 

\subsection{Semi-analytical modelling}
For the ICM density distribution, we adopt the  phenomenological model
of \cite{2013arXiv1311.4793Z}, which is based on gas profiles obtained
in X-rays \citep{2008A&A...487..431C}  and on an observational correlation between gas fraction and
mass of the clusters \citep{2009ApJ...693.1142S}. This method allows a gas density to be
assigned to 
any galaxy cluster using its mass alone, in such a way that the observed X-ray and Sunyaev-Zel'dovich 
scaling relations are correctly reproduced. 

For the CR spatial and spectral distribution, we adopt the hadronic model proposed in~\cite{2012arXiv1207.6410Z}, 
which extends the semi-analytical model of \cite{2010MNRAS.409..449P}. The latter provides a scaling of the 
CR distribution with  the cluster mass, while \cite{2012arXiv1207.6410Z} introduced an effective parameterisation 
on the CR spatial distribution $\rho_{\rm{CR}}$ to account for CR transport phenomena. 
In all the models analysed in this section, we assume the proton spectral shape 
as in~\cite{2010MNRAS.409..449P} where a universal CR spectrum is found amongst the simulated galaxy clusters. 
We rely on Equations~(\ref{eq:TOTgamma}) and (\ref{eq:Nradio}) with $L_{\gamma}(M_{500},z)$ and $L_{1.4}(M_{500},z)$ 
calculated by using Equations~(\ref{eq:Lgamma}) and (\ref{eq:Lradio}),  with $\rho_{\rmn{ICM}}$ and $\rho_{\rmn{CR}}$
from the \cite{2013arXiv1311.4793Z,2012arXiv1207.6410Z} models, including redshift evolution.

The cluster population is divided into 50$\%$ cool-core and 50$\%$ non-cool-core clusters (as from
observations; see, e.g.,~\citealp{2007A&A...466..805C}) with different parameterisation of the ICM and CR profiles. 
Cool-core clusters are relaxed objects, so CRs could stream out of the core, 
creating flat CR profiles. Non-cool-core clusters are more turbulent objects that should cause 
CRs to advect with the gas and create centrally peaked CR profiles. The difference between 
cool-core and non-cool-core clusters is modelled through the parameter $\gamma_{\rmn{tu}} = \tau_{\rmn{st}}/\tau_{\rmn{tu}}$,
i.e., the ratio between the characteristic time scale of streaming and that of turbulence. This parameter ranges 
from 100 for highly turbulent cluster and centrally peaked CR distributions to 1 for relaxed clusters and flat 
distributions as CRs move towards the outskirts \citep{2012arXiv1207.6410Z}. 
Here, we assume $\gamma_{\rmn{tu}} = 3$  and 1 for loud and quiet cool-core clusters, 
and $\gamma_{\rmn{tu}} = 60$ and 1 for loud and quiet non-cool-core clusters, respectively.

The magnetic field is assumed to radially scale as the gas density:
\begin{equation}
B(r) =
 B_0\,\left(\frac{\rho_{\rmn{ICM}}(r)}{\rho_{\rmn{ICM}}(0)}\right)^{\alpha_{\rm
 B}},
\label{eq:B}
\end{equation}
where $B_0$ is the central magnetic field, and $\alpha_{\rmn{B}} = 0.5$ describes the
declining rate of the magnetic field strength towards the cluster outskirts
\citep[][and references therein]{2008A&A...482L..13D,2010A&A...513A..30B, 2011A&A...529A..13K}.
In particular, for quiet clusters, we adopt a central magnetic field $B_0$ of 4~$\mu$G (7.5~$\mu$G) 
for non-cool-core (cool-core) clusters, while we
choose 6~$\mu$G (10~$\mu$G), to account for the potential turbulent dynamo in loud objects.

\subsection{Results: gamma-ray and neutrino backgrounds}
\begin{figure*}[hbt!]
\centering
\includegraphics[width=0.457\textwidth]{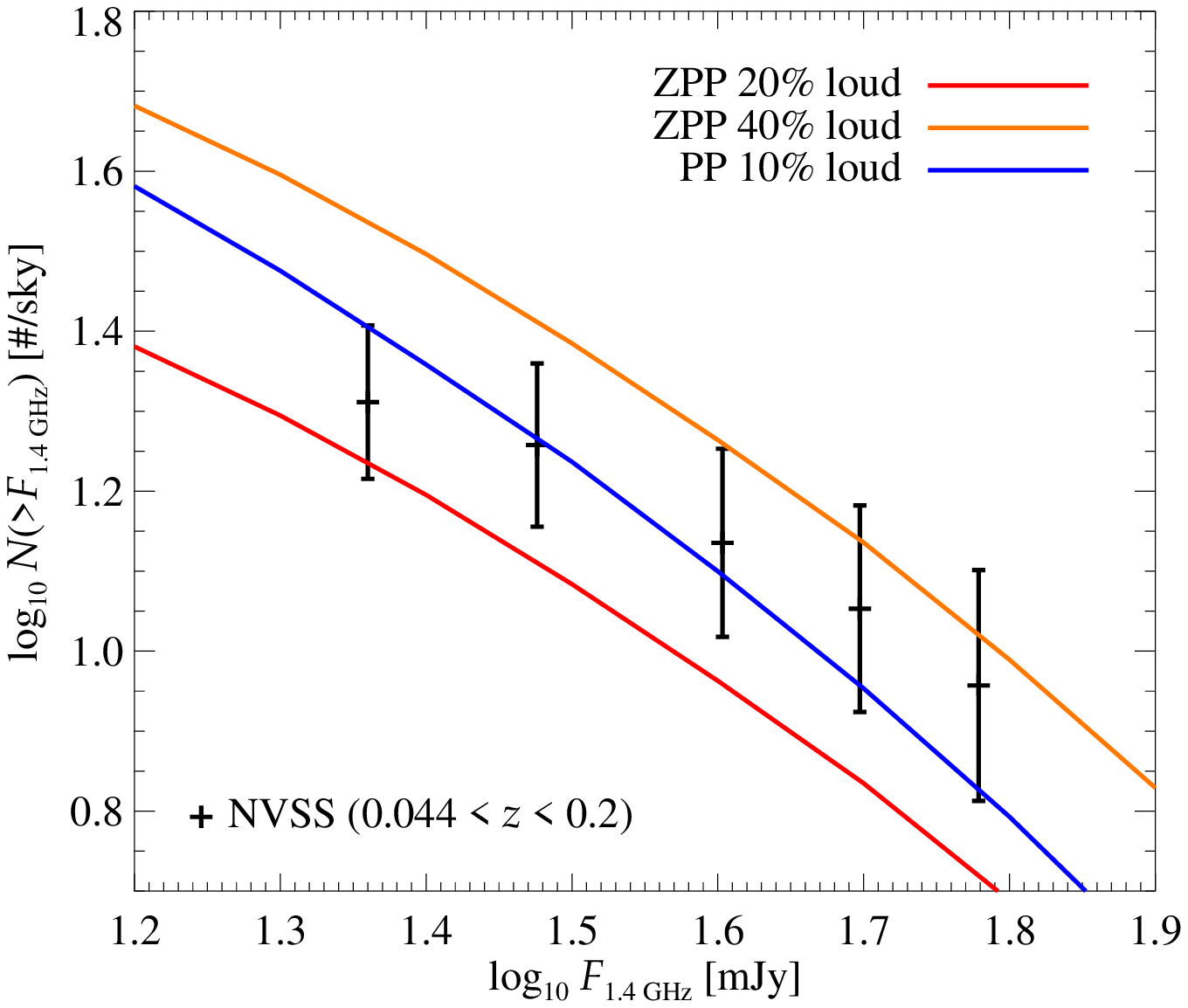}
\includegraphics[width=0.536\textwidth]{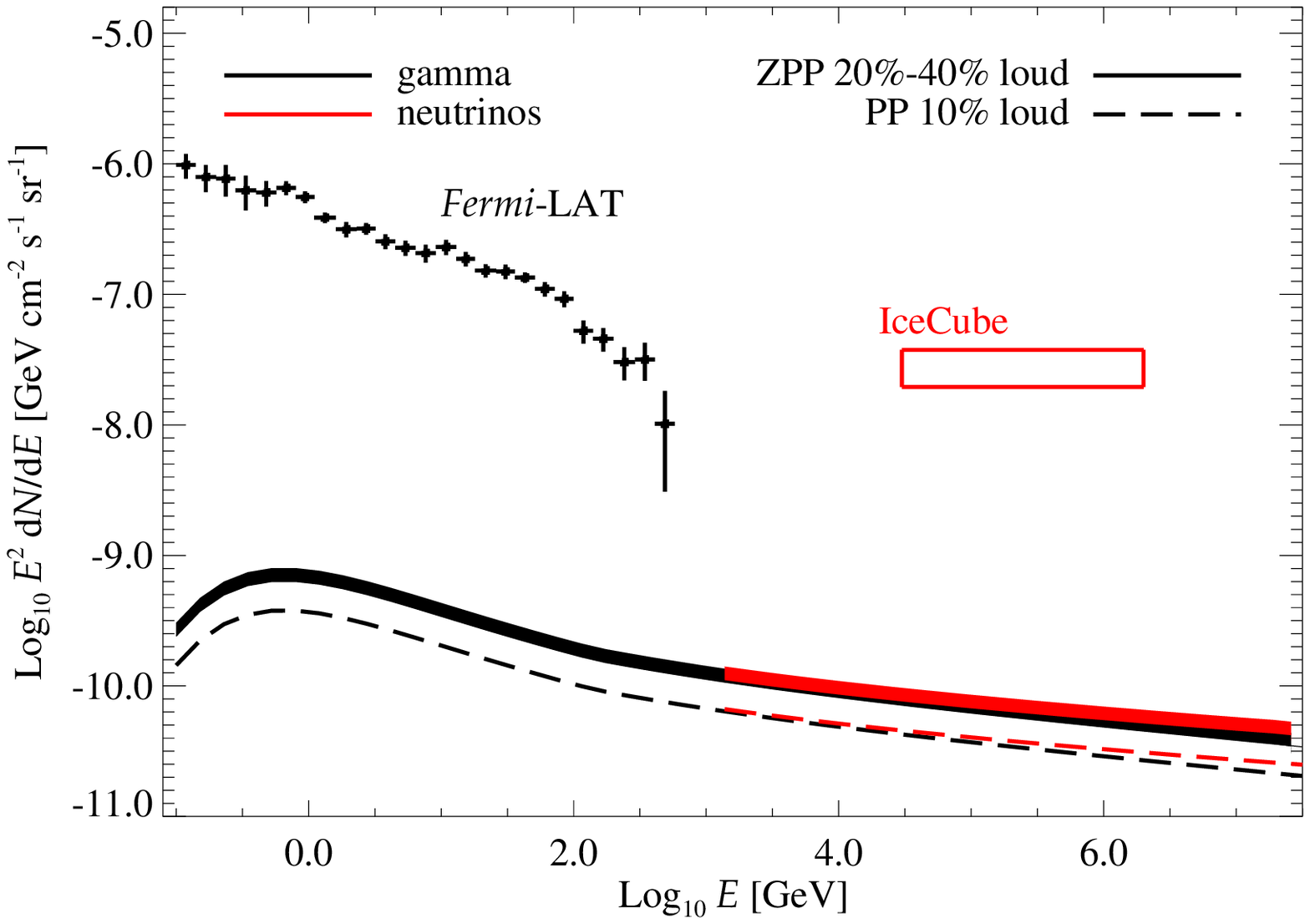}
\caption{
Radio counts due to synchrotron emission of secondary electrons as from the semi-analytical 
model of \cite{2012arXiv1207.6410Z} (ZPP in the plots) on the left, and total gamma-ray and neutrino intensities
on the right. For comparison, we plot  the {\it Fermi}~\citep{2014arXiv1410.3696T} 
and IceCube data~\citep{Aartsen:2014gkd} in the panels on the right. The neutrino intensity is meant for all flavours. We show the cases 
of the model applied to a mass function for 20\% and 40\% loud clusters, and additionally for 10\% 
loud clusters with parameters as in the \cite{2010MNRAS.409..449P} model with a maximum CR proton acceleration efficiency 
of 15\% (PP in the plots). According to this semi-analytical model, galaxy clusters contribute $<1$\% to 
the diffuse gamma-ray and neutrino backgrounds.
}
\label{fig:results}
\end{figure*}

The model in \cite{2012arXiv1207.6410Z} (ZPP in tables and figures) reproduces the observed radio-to-X-ray and radio-to-Sunyaev-Zel'dovich scaling relations
of galaxy clusters and respects current gamma-ray upper limits.\footnote{The parameters for the corresponding 
$L_{\gamma} (100\,\rmn{MeV})$--$M_{500}$ scaling relation at $z = 0$ are $P_1 = 21.68$ and $P_2 = 1.62$ for 
non-cool-core clusters, and $P_1 = 22.41$ and $P_2 = 1.57$ for cool-core clusters. This translates in Coma-like and 
Perseus-like fluxes, for $\alpha_{\rmn{p}}=2.2$, of $F(>100\rmn{MeV}) = 1.6\times10^{-9}$ and $F(>1\,\rmn{TeV}) = 7.6\times10^{-14}$~cm$^{-2}$~s$^{-1}$, 
respectively, below the current upper limits.} In the left-hand panel of Figure~\ref{fig:results}, we show the resulting
radio counts for a fraction of 20\% and 40\% loud clusters. We find that the latter case should be considered extreme because
hadronic interactions are known not to be uniquely responsible for the observed radio emission in clusters.
Table~\ref{tab:LM_trials_1} shows the corresponding total gamma-ray and neutrino fluxes.

Figure~\ref{fig:results} (left panel) also shows the radio counts obtained by adopting 10\% loud clusters
with parameters corresponding to the model in \cite{2010MNRAS.409..449P} (PP in tables and figures) with a 
maximum CR proton acceleration efficiency scaled down to 15\% with respect to the originally 
assumed 50\% in order to obey current gamma-ray constraints \citep{2014MNRAS.440..663Z,2013arXiv1308.5654T}. 
For the remaining 90\% quiet fraction, the parameters of the previous model are assumed.
The right-hand panel of Figure~\ref{fig:results} shows the corresponding
total gamma-ray and neutrino intensities compared with the data from
\emph{Fermi} and IceCube. We conclude that galaxy clusters contribute less than 1\% to the diffuse gamma-ray and 
neutrino backgrounds. 

The results reported in this section are more realistic than the ones shown in Section~\ref{sec:3}. 
However, we underline as  the semi-analytical model adopted here is based on the hypothesis that CRs are 
accelerated at structure formation shocks, while no assumption on the
CR sources is made in the phenomenological approach of Section~\ref{sec:3}.

\begin{table}[t]
\begin{center}
  \caption{\label{tab:LM_trials_1} Total gamma-ray and neutrino fluxes for the semi-analytical model.
    }
\resizebox{0.48\textwidth}{!}{
\begin{tabular}{lcc}
\hline\hline
\phantom{\Big|}
Model & $I_{\gamma}$ (100~MeV) & $I_{\nu}$ (250~TeV) \\
\hline\\[-0.5em]
ZPP 40\% & $3.0$ & $1.3$ \\
ZPP 20\% & $2.4$ & $1.0$ \\
ZPP 20\% $z_{2}=0.6$ & $2.0$ & $0.9$\\
ZPP 20\% $M_{500,\rmn{lim}}=10^{13} h^{-1} M_\odot$ & $6.2$ & $2.3$\\
ZPP 20\% $Planck$ & $4.2$ & $1.7$\\
PP 10\% & $1.5$ & $0.6$\\
\hline
\end{tabular}
}
\end{center}
\small{{\bf Note.} Total gamma-ray and neutrino flux at 100~MeV and 250~TeV for the 
semi-analytical model in units of $10^{-8}$ and $10^{-21}$~cm$^{-2}$~s$^{-1}$~GeV$^{-1}$~sr$^{-1}$, respectively.
} 
\end{table}

\subsection{Results: dependence on cosmology and lower mass bound}
\label{sec:4.3}
To test the robustness of our results, we computed the gamma-ray and neutrino backgrounds in the case of 20\% loud clusters, first  
extending the integration down to lower masses ($M_{500,\,\rmn{lim}}=10^{13} h^{-1} M_\odot$) and then adopting the most 
recent \emph{Planck} results for the cosmological parameters. 

\begin{figure*}[hbt!]
\centering
\includegraphics[width=0.495\textwidth]{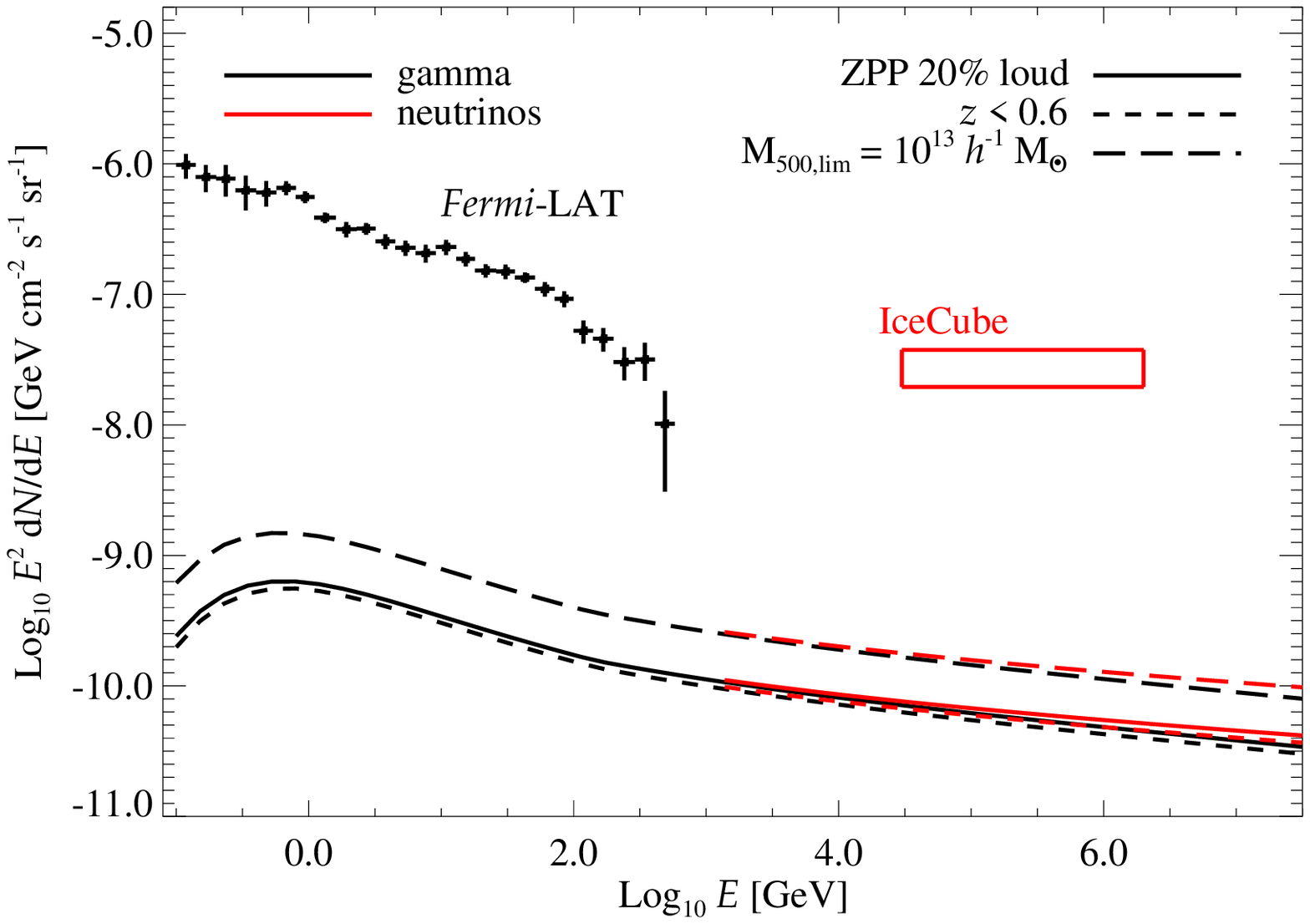}
\includegraphics[width=0.495\textwidth]{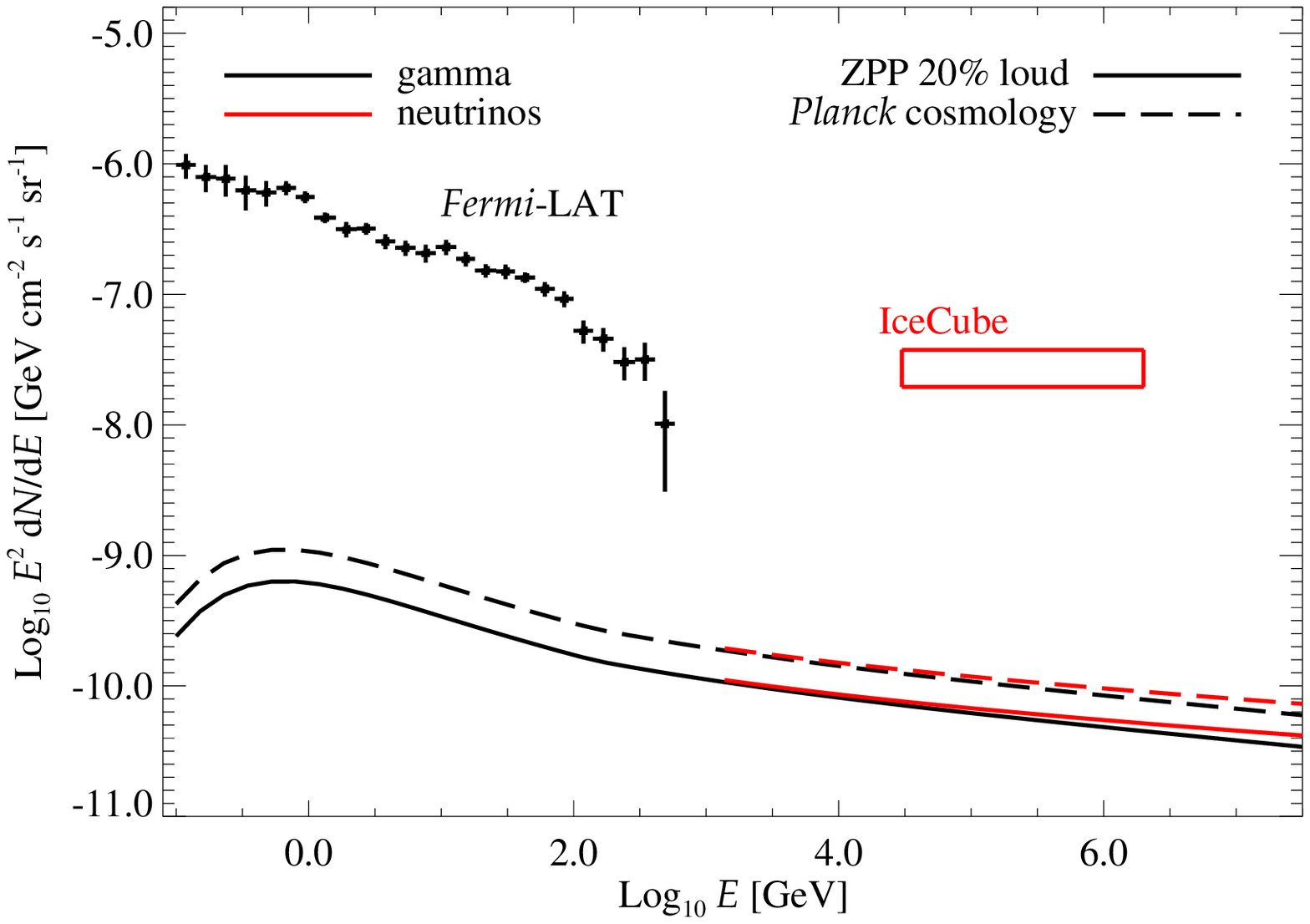}
\caption{Same as in the left panel of Figure~\ref{fig:results} for 20\% loud clusters. The left panel shows the comparison with the previous model
with one obtained adopting $z_2 = 0.6$, and one with a lower mass integration limit, of $10^{13}$~$h^{-1}$~$M_{\odot}$. The right panel
shows the comparison with the model in Figure~\ref{fig:results} and one obtained using the $\emph{Planck}$ cosmological data.
}
\label{fig:results_1}
\end{figure*}

The left-hand panel of Figure~\ref{fig:results_1} shows the gamma-ray and neutrino backgrounds for the same case as shown in 
Figure~\ref{fig:results} for 20\% loud clusters with $M_{500,\,\rmn{lim}}=10^{13.8} h^{-1} M_\odot$ and for 
$M_{500,\,\rmn{lim}}=10^{13} h^{-1} M_\odot$. In the latter case, the gamma-ray and neutrino diffuse fluxes are significantly
higher, while still representing less than 1\% of the observational data. At the same time, the radio counts are exactly the same
as in Figure~\ref{fig:results} since these are due to the higher mass objects.
We additionally show the case with $M_{500,\,\rmn{lim}}=10^{13.8} h^{-1} M_\odot$ integrated up to 
$z_{2} = 0.6$. As anticipated in Section~\ref{sec:2}, low-redshift objects represent the dominant contribution to the diffuse fluxes, 
because by adopting $z_{2} = 0.6$, we obtain 82\% of the total flux.

The right-hand panel of Figure~\ref{fig:results_1} shows gamma-ray and neutrino backgrounds for the same case as shown in 
Figure~\ref{fig:results} for 20\% loud clusters and obtained by adopting the cosmological parameters determined by the 
\emph{Planck} satellite~\citep{2013arXiv1303.5076P}, i.e., $H_0 = 67.3$~km~s$^{-1}$~Mpc$^{-1}$, 
$\Omega_\rmn{m}=0.32$, $\Omega_{\Lambda}=0.68$, and the corresponding mass function. The \emph{Planck} cosmology results in 
an overall larger number of structures, as is clear in Figure~\ref{fig:massfun}, therefore increasing both the total radio counts (not shown,
but still below the 40\% loud case of Figure~\ref{fig:results}) and the total gamma-ray and neutrino fluxes. As shown 
in Figure~\ref{fig:results_1}, the contribution to the extragalactic gamma-ray and neutrino background is at any rate lower than $1\%$.

We note that the changes in $M_{500,\,\rmn{lim}}$ and in the cosmological parameters would affect the gamma-ray and neutrino diffuse fluxes 
obtained with the phenomenological approach in Section~\ref{sec:3.2} approximately in the same way, i.e., they would 
increase by a factor of around 2, as can be seen from Table~\ref{tab:LM_trials_1}.

\section{Comparison with stacking limits by IceCube and future detection prospects}
\label{sec:4a}
Recently, \cite{2014ApJ...796..109A} have presented an all-sky point- and extended-source search with one-year IceCube data.
In particular, they provide upper limits on a stacked sample of nearby galaxy clusters, namely Virgo, Centaurus, Perseus, Coma, and Ophiuchus,
following predictions provided by \cite{2008ApJ...689L.105M}. Here, we focus on comparing with their ``model B,'' where CR protons are supposed
to be uniformly distributed within the cluster virial radius, and with their ``isobaric model," where CRs are assumed to be distributed like the ICM in the clusters. 

Following \cite{2011ApJ...732...18A}, we estimate the summed output of the five aforementioned clusters to be $I_{\nu}(250~\rmn{TeV}) = 1.1 \times 10^{-20}$
and $1.6 \times 10^{-20}$~cm$^{-2}$~s$^{-1}$~GeV$^{-1}$ for Model B and the isobaric model, respectively. The latest results by \cite{2014ApJ...796..109A}
provide the following upper limits $I_{\nu,\,\rmn{UL}}(250~\rmn{TeV}) = 6.9 \times 10^{-20}$  for 
Model B and $7.7 \times 10^{-20}$~cm$^{-2}$~s$^{-1}$~GeV$^{-1}$  for the isobaric model.

In Table~\ref{tab:cluster_sample}, we provide the maximum allowed neutrino flux for the same five clusters by adopting the phenomenological 
luminosity-mass relations obtained in Section~\ref{sec:3}. We use the same mass of these clusters as from the literature\footnote{The mass
$M_{500}$ for Centaurus, Perseus, Coma, and Ophiuchus is taken from \cite{2002ApJ...567..716R}, while for Virgo it is derived from 
\cite{2011arXiv1105.3240P}.} in order to apply our $L_{\gamma} (100\,\rmn{MeV})$--$M_{500}$ relation, and\ they should therefore be 
considered indicative numbers, as in Section~\ref{sec:3} for Coma and Perseus. We use $\alpha_\rmn{p} = 2, 2.2$, and $2.4$, omitting the 
extreme case of $1.5$, and always refer to the case with $B = 1$~$\mu$G. For $\alpha_\rmn{p} = 2$, we adopt the $P_1$ value 
for 30\% loud clusters. See Table~\ref{tab:LM_trials} for more details.

The upper limits $I_{\nu,\,\rmn{UL}}(250~\rmn{TeV})$ for this stacked sample of clusters by \cite{2014ApJ...796..109A} are obtained
by assuming a spectral index of $\approx 2.15$, so we can compare with our results for $\alpha_\rmn{p} = 2.2$. From 
Table~~\ref{tab:cluster_sample}, we can see that the corresponding IceCube upper limits are just a factor of $1.3-1.5$ above
the maximum allowed flux for the stacked sample. When $\alpha_\rmn{p} = 2.4$, the maximum allowed flux for the stacked 
sample is one order of magnitude lower, while for $\alpha_\rmn{p} = 2$ it is one order of magnitude higher, with respect to 
$\alpha_\rmn{p} = 2.2$. We can conclude that, while special care should be used in considering the profile and extension
of the possible signal, IceCube should be able to put constraints on our most optimistic case with $\alpha_\rmn{p} = 2$ 
and on the $\alpha_\rmn{p} = 2.2$ case in the very near future, while the case with $\alpha_\rmn{p} = 2.4$ is
much harder to achieve.

We underline that the fluxes presented in this section for Virgo, Centaurus, Perseus, Coma, and Ophiuchus
are quite optimistic for representing the maximum allowed by our phenomenological approach. For example, we know 
that the fluxes of Virgo, Centaurus, and Ophiuchus should lie significantly below the loud part of the luminosity-mass
relation owing to the lack of diffuse radio emission in Virgo and Centaurus, and to the very low surface-brightness
radio emission observed in Ophiuchus (see, e.g., \citealp{2012arXiv1207.6410Z}), pushing also the possible
hadronic-induced gamma-ray and neutrino fluxes to lower levels. Any realistic modelling of these objects 
should consider this evidence carefully. In fact, the stacked signal from the five nearby clusters
presented in this section already significantly overshoots the total signal obtained with the more realistic 
modelling of the CR proton population in clusters performed in Section~\ref{sec:4} with our semi-analytical 
approach.

\begin{table}[t]
\begin{center}
  \caption{\label{tab:cluster_sample} Maximum allowed neutrino flux from nearby clusters at 250~TeV.
    }
\begin{tabular}{lcccc}
\hline\hline
\phantom{\Big|}
Cluster & & $\alpha_\mathrm{p} = 2$ & $\alpha_\mathrm{p} = 2.2$ & $\alpha_\mathrm{p} = 2.4$ \\
\hline\\[-0.5em]
Virgo & $\le$ & $3.2 \times 10^{-19}$ & $4.0 \times 10^{-20}$ & $3.4 \times 10^{-21}$\\
Centaurus & $\le$ & $7.3 \times 10^{-21}$ & $9.1 \times 10^{-22}$ & $7.7 \times 10^{-23}$\\
Perseus & $\le$ & $1.8 \times 10^{-20}$ & $2.3 \times 10^{-21}$ & $1.9 \times 10^{-22}$\\
Coma & $\le$ & $2.8 \times 10^{-20}$ & $3.5 \times 10^{-21}$ & $2.9 \times 10^{-22}$\\
Ophiuchus & $\le$ & $4.5 \times 10^{-20}$ & $5.6 \times 10^{-21}$ & $4.7 \times 10^{-22}$\\
\hline\\[-0.5em]
Sum & $\le$ & $4.2 \times 10^{-19}$ & $5.2 \times 10^{-20}$ & $4.4 \times 10^{-21}$ \\
\hline
\end{tabular}
\end{center}
\small{{\bf Note.} Maximum allowed neutrino flux at 250~TeV in units of cm$^{-2}$~s$^{-1}$~GeV$^{-1}$.
Numbers were obtained assuming the phenomenological luminosity-mass relations of Section~\ref{sec:3}. 
All cases refer to $B = 1$~$\mu$G; $\alpha_\mathrm{p} = 2$ refers to 30\% loud clusters,
our most optimistic while still realistic case; and the cases of $\alpha_\mathrm{p} = 2.2$
and $2.4$ refer to 100\% loud clusters (see Table~\ref{tab:LM_trials}).
} 
\end{table}

\section{Proton-photon interactions in galaxy clusters}
\label{sec:5}
Besides interacting with the ICM, relativistic protons in clusters of galaxies can also interact with the ambient photon fields. The two main interaction processes are electron--positron pair production ($p + \gamma \rightarrow p + e^+ + e^-$) and photomeson production. (Close to the threshold, the dominant contribution comes from the resonant channel: $p + \gamma \rightarrow \Delta^{+} \rightarrow p + \pi^0$ or $n + \pi^-$.) Both photons and neutrinos are expected in photomeson production owing to the decay of neutral and charged pions, respectively \citep{2008PhRvD..78c4013K}. Thus, this is another channel to be investigated for assessing the contribution of clusters of galaxies to the diffuse neutrino flux observed by IceCube.

The process of photomeson production has a kinematic threshold and takes
place when the energy of the photon in the rest frame of the proton
exceeds $E_{\rm thr} \simeq 145$~MeV
\citep[see, e.g.,][]{2008PhRvD..78c4013K}. The most prominent radiation field
in clusters of galaxies is the CMB \citep[e.g.,][]{2011arXiv1105.3240P}, whose
photons have a typical energy of $E_{\rm CMB} \approx 7 \times
10^{-4}$~eV. The threshold energy for a proton to produce a meson is
$E_{\rm p,thr} = E^2_{\rm thr}/2 E_{\rm CMB} \approx 10^{20}$~eV, but in
fact protons with slightly smaller energy can also interact with the
high-energy tail of the black body radiation
\citep{1966PhRvL..16..748G,1966JETPL...4...78Z}. Thus, one can conclude
that proton-photon interactions in clusters of galaxies can contribute to the high-energy neutrino background only if
protons with energy in excess of several $10^{19}$~eV are present in the ICM.

Accretion shocks around clusters of galaxies have been proposed as
the sites of the acceleration of ultrahigh-energy CRs, the main
reason being that their very large size (Mpc scale) would allow the
acceleration and confinement of protons of ultrahigh energies
\cite[e.g.,][]{1995ApJ...454...60N}. An estimate of the maximum energy
achievable by protons at cluster accretion shocks can be obtained by
equating the acceleration time, computed in the framework of diffusive shock acceleration, to the energy loss time due to proton-photon interactions. Accurate calculations have shown that the maximum energy of protons is determined by the energy losses due to electron--positron pair production and that for the most optimistic assumptions it ranges from a few $10^{18}$~eV to a few $10^{19}$~eV \citep{2011A&A...536A..56V}. Because they are cooled mainly by pair production, protons are thus not expected to produce any appreciable flux of neutrinos through the proton-photon interaction channel. Heavy nuclei, such as iron, can be accelerated up to $\approx 10^{20}$~eV at cluster accretion shocks \citep[e.g.,][]{2009A&A...502..803A,2011A&A...536A..56V}. However, iron cools mainly by photodisintegration in a soft photon field, and in this case the neutrino yield is very suppressed compared to the case of photomeson production \citep{2009ApJ...707..370K}.

Another possible scenario for the production of neutrinos in the ICM would be to assume that clusters contain sources of ultrahigh-energy CRs. This would lead to two advantages. First of all, the infrared photon background in the cluster core would be enhanced with respect to the cosmological one thanks to the contribution from the galaxies in the cluster \citep{2005ARA&A..43..727L,2011arXiv1105.3240P}. Second, the turbulent magnetic field present within the ICM would partially confine ultrahigh-energy protons, enhancing the probability of interaction. These two facts would increase the expected neutrino flux from proton-photon interactions \citep[e.g.,][]{2006PhRvD..73d3004D,2009ApJ...707..370K}.
However, the source of ultrahigh-energy CRs will have to be located in the centre of the cluster, where the infrared photon background is enhanced and the confinement of protons is more effective (thanks to a larger magnetic field). As pointed out in \citet{2009ApJ...707..370K}, the high gas density in the core of clusters would also enhance the probability of proton--proton interactions, which would dominate the neutrino production below energies of $\approx 10^{18}$~eV.

Finally, it has to be noticed that the expected spectrum of neutrinos from photopion production interactions is significantly harder than $E^{-2}$ 
below the energy threshold, which is at odds with the evidence for a spectral index softer than two revealed by IceCube \citep{Murase:2013rfa,2014PhRvD..89l3005B}.
This implies that proton--photon interactions make a negligible contribution to the neutrino flux in the energy domain of the IceCube neutrinos.

\section{Contribution to the small-scale anisotropies of the gamma-ray
 background}
\label{sec:6}

Recently, \cite{2012PhRvD..85h3007A} has analysed the anisotropies in the EGB
and found an excess in its angular power spectrum over what is expected
with a completely diffuse source distribution on multipole ranges $155
\le \ell \le 504$ (corresponding to $\lesssim 2\degr$ angular scales).
For the first time, this  has shown that a major fraction of the EGB is made
by discrete sources, and, in fact, \cite{2012PhRvD..86f3004C} point out
that the measured level of anisotropies is consistent with predictions
for gamma-ray blazars \citep{Ando:2006cr}.
They also obtained the upper limit on the angular power spectrum as
$C_\ell < 3.3 \times 10^{-18} \ \mathrm{(cm^{-2}~ s^{-1} ~ sr^{-1})^2 ~
sr}$ for $155 \le \ell \le 504$ and $E = 1$--10~GeV on other source
components, after subtracting the main blazar contribution.
Even though clusters are not the dominant contributors to the
isotropic component of the diffuse gamma-ray and neutrino backgrounds (as
shown in the previous sections), they may make substantial contributions
to the EGB {\it \emph{anisotropies}}.
In particular, since there are relatively fewer than
other astrophysical sources, such as star-forming galaxies, the cluster
component in the EGB should be more anisotropic.
To this end, we estimate the cluster contribution to
the EGB anisotropies in this section and compare it to the {\it Fermi} data at
sub-degree angular scales.

\begin{figure}[hbt!]
\centering
\includegraphics[width=0.5\textwidth]{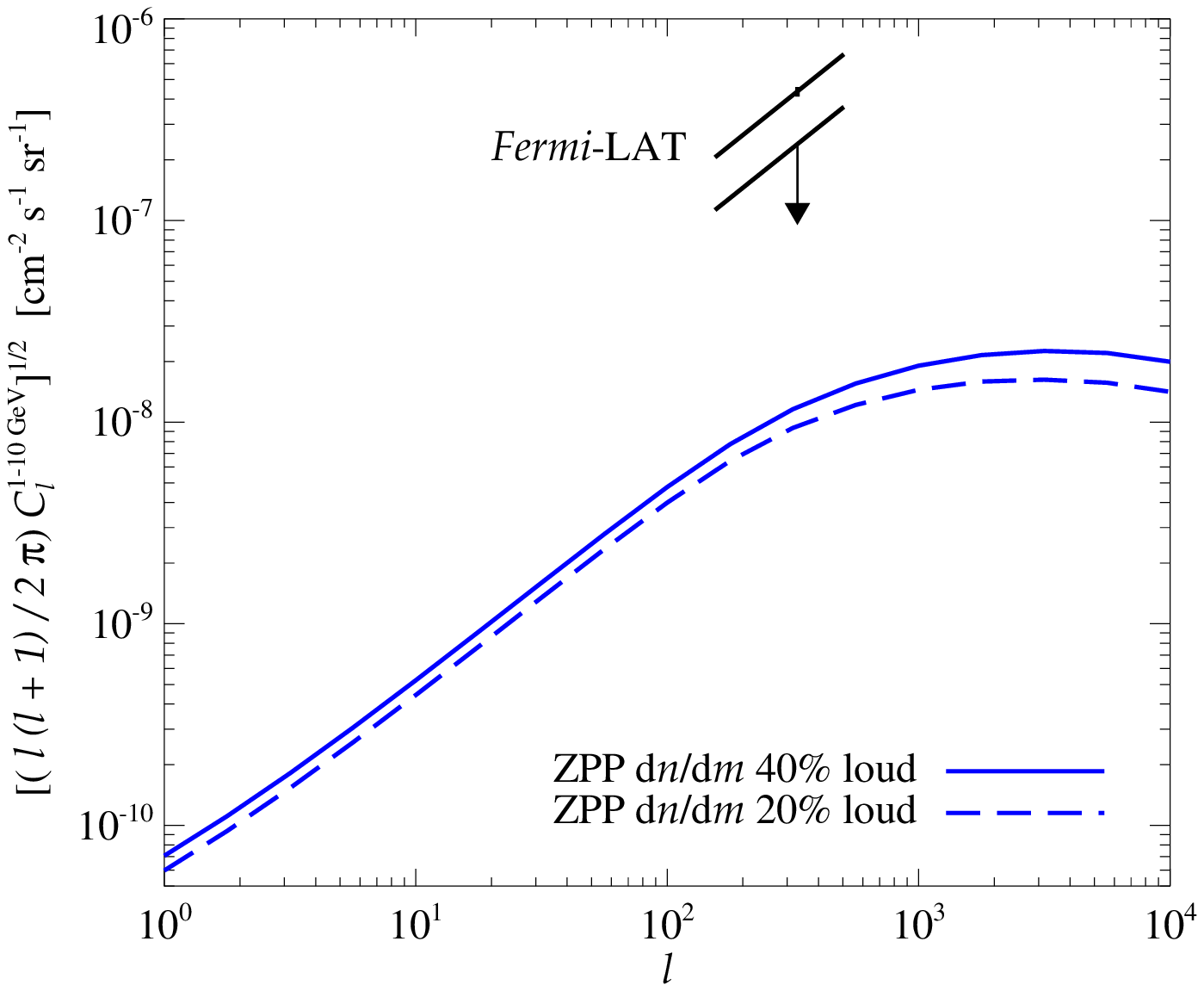}
\caption{Gamma-ray angular power spectrum for emission resulting from proton-proton interactions in galaxy clusters
in the energy range $1-10$~GeV. We show the result for the semi-analytical model of Section~\ref{sec:4} for 20\% and 40\% 
loud clusters. We plot  the EGB anisotropy measured by \emph{Fermi} \citep{2012PhRvD..85h3007A} for comparison, which is 
explained by unresolved blazars, and the upper limits obtained once the blazar component is subtracted \citep{2012PhRvD..86f3004C}.
We plot the square root of $\ell(\ell+1)C_{\ell}/2\pi$, which implies that the shown quantity is directly proportional to an increase in
intensity.
}
\label{fig:results_ps}
\end{figure}

The angular power spectrum coming from proton-proton interactions in
galaxy clusters can be calculated as
follows \citep[e.g.,][]{Ando:2006cr}:
\begin{equation}
C_{\ell} = \int \frac{\rmn{d}\chi}{\chi^2} \, W_{\gamma}^2(E\,[1+z],z)\,P_{\rm{C}}\left( k=\frac{\ell}{\chi},\chi \right) \, ,
\label{eq:PS}
\end{equation}
where $\chi$ is the comoving distance (we use the same redshift range as in previous sections), $W_\gamma= (1+z)^3 A_{\gamma}(E\,[1+z])/4\pi$ 
is the so-called window function, and $P_{\rmn{C}}(k,\chi)$ is the power
spectrum for the cluster gamma-ray emission. The last can be divided into one- 
and two-halo terms, $P_{\rmn{C}} = P_{\rmn{C}}^{\rm{1h}} +
P_{\rmn{C}}^{\rm{2h}}$, which we express as \citep[e.g.,][]{Komatsu:2002wc, Ando:2006cr}
\begin{eqnarray}
P_{\rmn{C}}^{\rm{1h}} & = & \int \rmn{d}M \frac{\rmn{d}n}{\rmn{d}M} \left[ \int 4\pi r^2 \rmn{d}r \rho_\rmn{CR}(r) \rho_\rmn{gas}(r) \frac{\sin(kr)}{kr}  \right]^2 \, , \\
P_{\rmn{C}}^{\rm{2h}} & = & \left[ \int \rmn{d}M \frac{\rmn{d}n}{\rmn{d}M} \, b(M,z ) \int 4\pi r^2 \rmn{d}r \rho_\rmn{CR}(r) \rho_\rmn{gas}(r) \frac{\sin(kr)}{kr} \right]^2  \nonumber \\
& & \times \, P_{\rmn{lin}} (k,\chi) \, ,
\end{eqnarray}
respectively, where the radial integration goes up to $R_{500}$. In the
two-halo term, we assume that the linear matter power spectrum $P_{\rmn{lin}}(k,\chi)$ is related 
to the cluster power spectrum via the linear bias $b(M,z)$ \citep{2010ApJ...724..878T}. We find that the one-halo term dominates the two-halo
term at all multipoles $\ell$.

In Figure~\ref{fig:results_ps}, we show the angular power spectrum for the semi-analytical models 
of Section~\ref{sec:4} for 20\% and 40\% loud clusters integrated in the energy bin from 1 to 10~GeV. 
We compare with the measurement on the EGB by \emph{Fermi}
\citep{2012PhRvD..85h3007A} and upper limits by \cite{2012PhRvD..86f3004C}.
We compare $C_\ell^{1/2}$ instead of $C_\ell$. This is because 
$C_\ell$ is a variance, so if each cluster is twice as bright, 
then $C_\ell$ becomes larger by a factor of 4. Therefore, taking the 
square-root will reflect the correct scaling with respect to the cluster 
contribution. Our prediction is about one order of magnitude less 
than the {\it Fermi} upper limit. This means that in scenarios where the total 
galaxy cluster intensity is much higher than in the models of
Section~\ref{sec:4}, as is potentially realised for some of the simple 
phenomenological models discussed in Section~\ref{sec:3}, the angular power
spectrum could be a powerful discriminator, as powerful as radio counts.
Additionally, there are other contributions to the EGB anisotropies
that would further increase the gamma-ray angular power spectrum, such as,
but not only, dark matter annihilation (e.g., \citealp{2006PhRvD..73b3521A,2013MNRAS.429.1529F,
2013PhRvD..87l3539A,2015arXiv150105464T}), further exacerbating
the possible tension with the upper limits by \cite{2012PhRvD..86f3004C}.

\section{Discussion and conclusions}
\label{sec:7}
In this work we estimated the contribution from hadronic proton-proton interactions in galaxy clusters to the total extragalactic gamma-ray
and neutrino fluxes, while including radio constraints for the first time. We modelled the cluster population by means of their mass function. 
Our approach makes use of a phenomenological luminosity-mass relation applied to all clusters, constructed by requiring radio counts
to be respected. We adopted four different proton spectral indices 
$\alpha_\rmn{p} = 1.5, 2, 2.2,$ and $2.4$, and three different magnetic field values $B>>B_\rmn{CMB}$, $B=1$~$\mu$G, and $B=0.5$~$\mu$G.
The last is meant to only be an illustrative case, because it contrasts with current estimates of magnetic fields in clusters.

Radio observations reveal that not all galaxy clusters host diffuse synchrotron radio emission, with upper limits about an order of 
magnitude below the loud state \citep{2009A&A...507..661B,2013arXiv1306.4379C}.  
For the sake of simplicity, we adopted 100\% loud clusters, leading to an optimistic estimation.  
However, we also discussed the case with 30\% loud clusters for $\alpha_\rmn{p} = 2$, corresponding to our most
optimistic case, according to recent estimates of the loud fraction. In our phenomenological model, the slope
of the luminosity-mass relation is fixed to 5/3, assuming that the hadronic-induced luminosity scales as the 
cluster thermal energy, and the redshift evolution was omitted for simplicity. We showed that our assumptions are robust, 
and we estimated that ignoring the redshift evolution results in only about a 20\% underestimation of
the radio counts and total high-energy fluxes.

By requiring all the current constraints to be respected from radio counts to gamma-ray upper limits on individual clusters, we showed that 
galaxy clusters can contribute at most up to $10\%$ of the total neutrino background for $\alpha_\rmn{p} = 2$, while contributing much less to the EGB. 
For $\alpha_\rmn{p} > 2$, the gamma-ray and neutrino backgrounds in all considered cases are $<1\%$ of the gamma-ray and neutrino fluxes measured 
by \emph{Fermi} \citep{2010PhRvL.104j1101A,2014arXiv1410.3696T} and IceCube \citep{Aartsen:2014gkd}, respectively. Only for the extreme case 
with $\alpha_\rmn{p} = 1.5$ is the neutrino flux of the same order of magnitude as the IceCube data; however, such a hard spectral shape contrasts 
with the most recent IceCube spectral fit of neutrino flux \citep{Aartsen:2014yta}. 

We also adopted a more refined approach that employs a semi-analytical model where the ICM density is constructed
from X-ray observations, and the CR spatial and spectral distribution is based on state-of-art hydrodynamic simulations \citep{2012arXiv1207.6410Z}. 
In this case, we divided the cluster population into cool-core/non-cool-core and loud/quiet subsamples, as suggested by 
observations, where the transition from the loud to the quiet state is achieved through a change in the CR propagation properties.
We find that galaxy clusters contribute to $<1\%$ to the EGB and to the neutrino flux measured by IceCube.
While this semi-analytical model is more realistic than the simplified phenomenological model discussed above, 
we assume in this case that CRs  are accelerated at structure formation shocks, while no assumption on the CR sources was
made in the phenomenological approach.

We then compared the flux of five nearby clusters - Virgo, Centaurus, Perseus, Coma and Ophiuchus - to 
recent results by IceCube \citep{2014ApJ...796..109A}. The IceCube upper limits are just a factor $1.5$ above 
our maximum allowed (stacked) flux for these objects for the case of $\alpha_\rmn{p} = 2.2$, which compares well 
to what is used in \cite{2014ApJ...796..109A}. We showed that, despite the small contribution to the
total neutrino flux, IceCube should be able to put constraints on our most optimistic case with $\alpha_\rmn{p} = 2$, 
and very soon in the case with $\alpha_\rmn{p} = 2.2$, using the stacked sample of nearby massive clusters.

We briefly also discussed the case of proton-photon interactions in galaxy cluster. We found that 
this channel gives a negligible contribution to the expected neutrino flux in the multi TeV--PeV energy domain.

While galaxy clusters represent a sub-dominant contribution to the EGB, they could 
substantially contribute to its anisotropy because they are fewer in number
than  other astrophysical sources and, therefore, are expected to  be
more anisotropic. For this reason, we computed the angular power spectrum for the considered 
semi-analytical models and showed that the amplitude of the angular fluctuations, 
represented by $C_\ell^{1/2}$, is about one order of magnitude below the {\it Fermi} upper limits.

We conclude that there is no realistic scenario in which galaxy clusters can contribute substantially
to either the EGB or the extragalactic neutrino flux, since the maximum  contribution is at most
10\% in the simple phenomenological modelling, while it is  less than 1\% in most cases and in the more realistic semi-analytical modelling.
We also proved that our conclusions are not significantly affected by our assumptions. 
Our results therefore put earlier works into prospective, which turned out to be overly optimistic in estimating the 
galaxy cluster contribution (e.g., \citealp{2000Natur.405..156L,2013PhRvD..88l1301M}).

We would like to conclude with a few additional comments on our assumptions. In our calculations, we omitted both a possible cut-off in the CR spectrum 
at high energies caused by protons that are no longer confined to the cluster and the absorption of high-energy gamma rays 
due to interactions with the extragalactic background light. The former implies larger high-energy neutrino fluxes, while the latter 
implies slightly optimistic gamma-ray fluxes. Additionally, we stress once more how requiring the synchrotron emission from secondary 
electrons not to overshoot radio counts also results in rather optimistic gamma-ray and neutrino fluxes. This is because
so-called giant radio haloes hosted in merging, non-cool-core clusters cannot be explained solely by hadronic emission 
\citep{2012arXiv1207.3025B,2012arXiv1207.6410Z}. Therefore, the secondary emission seems to represent only a fraction of the total 
observed radio emission. 

As a final note on the semi-analytical modelling, we underline that the transition 
from the loud to the quiet state in the galaxy cluster population is not achieved in the \emph{\emph{classical}} hadronic model, meaning 
that it predicts that all clusters should have the same level of secondary emission. This clearly contradicts observations and 
represents one of the problems with the hadronic scenario (see \citealp{2011A&A...527A..99E} for a discussion). The only 
mechanism that has been proposed so far to solve this problem is to vary CR propagation properties (see, e.g., \citealp{2013arXiv1303.4746W}), 
which was also adopted in our semi-analytical approach through the \cite{2012arXiv1207.6410Z} model. We note, however, that it is still being 
debated whether the conditions for CR diffusion can be reached in the ICM. In the worst-case scenario, the secondary 
electrons produced in proton-proton collisions in clusters would only be seed electrons for subsequent  turbulent re-acceleration 
(see, e.g., \citealp{2010arXiv1008.0184B,2012arXiv1207.3025B}). This would imply a much lower secondary 
emission only at the level of the quiet state. If this turns out to be the case, the total gamma-ray and neutrino 
fluxes from galaxy clusters should be even lower than what we have estimated here.\\

\begin{acknowledgements}
We thank the anonymous referee for useful comments.
We thank Denis Allard, Rossella Cassano, and Kohta Murase for useful discussions. 
This work was supported by the Netherlands Organisation for Scientific 
Research (NWO) through a Vidi grant (SA, IT, and FZ) and a PHC Van Gogh grant (SG).
\end{acknowledgements}

\bibliographystyle{aa}
\bibliography{bib_file}

\label{lastpage}

\end{document}